\documentclass[onecolumn,usenatbib]{mn2e}
\usepackage{graphicx}
\usepackage{natbib}
\newif\ifAMStwofonts

\newcommand{\etal}{{et~al.}}

\newcommand{\fastica}{\sc{FastICA}}
\newcommand{\lightica}{\sc{lightICA}}
\newcommand{\lsim}{\,\lower2truept\hbox{${<\atop\hbox{\raise4truept\hbox{$\sim$}}}$}\,}
\newcommand{\gsim}{\,\lower2truept\hbox{${>\atop\hbox{\raise4truept\hbox{$\sim$}}}$}\,}

\def\aa{{\sl Astron.\ \&\ Astrophys.\ }}
\def\aas{{\sl Astron. \& Astrophys.\ Suppl.\ }}
\def\apj{{\sl Astrophys.\ J.\ }}

\def\apjs{{\sl Astrophys.\ J.\ Supp.\ }}

\def\ieeespl{{\sl IEEE\ Signal\ Processing\ Lett.\ }}

\def\jiee{{\sl J.\ Inst.\ Elect.\ Eng.\ }}

\def\mnras{{\sl MNRAS\ }}
\def\n{{\sl Nature\ }}

\def\nar{{\sl New\ Astron.\ Rev.\ }}

\def\prd{{\sl Phys.\ Rev.\ D\ }}

\def\pieee{{\sl Proc.\ IEEE\ }}

\def\s{{\sl Science\ }}

\title[Separating polarized cosmological and galactic emissions]
{Separating polarized cosmological and galactic emissions
for CMB B-mode polarization experiments}

\author[Federico Stivoli, Carlo Baccigalupi, Davide Maino, Radek Stompor]
{Federico Stivoli$^{1}$, Carlo Baccigalupi$^{1,2,3,4}$, Davide Maino$^{5}$, Radek Stompor$^{6,7,8}$\\$^{1}$ SISSA/ISAS, Astrophysics Sector, Via Beirut, 4,
I-34014 Trieste, Italy\\
$^{2}$ Institut f$\ddot{\it u}$r Theoretische Astrophysik,
Universit$\ddot{\it a}$t Heidelberg, Albert-berle-Strasse 2, D-69120
Heidelberg, Germany\\
$^{3}$ INFN, Sezione di Trieste, Via Valerio 2,
I-34014 Trieste, Italy\\
$^{4}$ Physics Division, Lawrence Berkeley National Laboratory, 1 
Cyclotron Road, Berkeley, CA 94720, USA\\
$^{5}$ Dipartimento di Fisica, Universit\`a di Milano,
Via Celoria 16, I-20133, Milano, Italy\\
$^{6}$ Computational Research Division, 
Lawrence Berkeley National Laboratory, 1 Cyclotron Road, Berkeley,
CA 94720, USA\\
$^{7}$ Space Sciences Laboratory, University of California,
Berkeley, CA 94720, USA\\
$^{8}$ Laboratoire Astroparticule et Cosmologie, Universit{\'e} Paris-7, Paris, France
}

\begin{document}

\maketitle

\label{firstpage}
\footnotetext{E-mail: stivoli@sissa.it}

\begin{abstract}
The detection and characterization of the 
$B$ mode of Cosmic Microwave Background
(CMB) polarization anisotropies will not be possible without 
a high precision removal of the foreground contamination present
in the microwave band. In this work we study the 
relevance of the component separation technique based on the Independent 
Component Analysis (ICA) for this purpose and investigate its performance
in the context of a limited sky coverage observation and from the viewpoint 
of our ability to differentiate between cosmological
models with different primordial B-mode content. \\
We focus on the low Galactic emission sky patch centered at 
40 degrees in right ascension and -45 in declination, 
corresponding to the target of several operating and planned CMB 
experiments and which, in many respects, adequately represents a typical 
``clean" high latitude sky.
We consider two fiducial observations, one operating at low (40, 90 GHz)
and one at high (150, 350 GHz) frequencies
and thus dominated by the synchrotron and thermal dust emission,
respectively. 
We use foreground templates simulated in accordance with 
the existing observations in the radio and 
infrared bands, as well as the Wilkinson Microwave Anisotropy Probe 
(WMAP) and Archeops data and model the CMB emission adopting the 
current best fit cosmological model, with an amplitude of primordial 
gravitational waves either set to zero or 10\%.
We use a parallel version of the {\fastica} code to explore a 
substantial parameter space including Gaussian pixel noise level, 
observed sky area and the amplitude of the foreground emission 
and employ large Monte Carlo simulations to
quantify errors and biases pertinent to  the  reconstruction for different 
choices of the parameter values.
We identify a large subspace 
of the parameter space for which the quality of the CMB reconstruction is excellent,
i.e., where the errors and biases introduced by the separation are found to be 
comparable or lower than the uncertainty due to the cosmic variance  and instrumental noise.
For both the cosmological models, with and without the primordial gravitational waves,
we find that {\fastica} performs extremely well even in the cases when the $B$ mode 
CMB signal is up to a few times weaker 
than the foreground contamination and the noise amplitude is comparable with the total 
CMB polarized emission.
In addition we discuss  
limiting cases of the noise and foreground amplitudes, for which the ICA approach fails. \\
Although our conclusions are limited by the absence of systematics in the simulated 
data, these results indicate that these component separation techniques could play 
a crucial role in the forthcoming experiments aiming at the detection of 
$B$ modes in the CMB polarization. 
\end{abstract}


\section{Introduction}
\label{introduction}

The main target of the planned probes for measuring the polarized 
component of the Cosmic Microwave Background (CMB) radiation is 
represented by the $B$ modes, also known as ``curl" component 
\citep{zaldarriaga_seljak_1997,kamionkowski_etal_1997}. The CMB $B$ mode signal
is known to be generated by primordial gravitational waves and the weak lensing 
due to structures forming in the Universe \citep{zaldarriaga_seljak_1998},
and thus contains unique information about the early Universe, and, potentially, physics of
high energies.\\
The $B$ signal in the CMB polarization is more than one order of 
magnitude smaller than the ``gradient" mode ($E$) coming from all kinds 
of cosmological perturbations, and about two orders of magnitude 
lower if compared with the total intensity anisotropies ($T$).
The CMB $E$ mode and the $TE$ cross-correlation, have been detected by 
the WMAP satellite \citep{page_etal_2006} as well as instruments operating on the ground 
\citep{kovac_etal_2002,readhead_etal_2004} and from balloons \citep{montroy_etal_2005}. 
No glimpse of the $B$ has been seen so far and it is apparent that its
detection will represent an experimental and data analysis challenge in terms of control 
and treatment of systematics and instrumental noise needed to attain the required precision.
An additional important limiting factor for these experiments is related to 
foreground emissions. In the frequency range going from 70 to 150 GHz the diffuse 
Galactic emission is known to be sub-dominant with respect to the total intensity 
CMB signal at medium and high Galactic latitudes. 
The knowledge of polarized foregrounds has been given a boost by the observations 
by Archeops \citep{benoit_etal_2004} and most importantly the WMAP three year data
\citep{page_etal_2006}. However, the foreground polarized pattern, expecially 
at high Galactic latitudes, remains quite unknown; these new findings confirm
the earlier guess that the weakness of the cosmological $B$ signal makes the foreground 
contamination, at least potentially and, more likely, certainly very important for the 
recovery of this component everywhere in the sky and at any frequency 
\citep{baccigalupi_2003}.

In this context, it is crucial to develop reliable data analysis techniques
and tools which are 
capable of cleaning the CMB emission from the foreground contamination, prior
to assessing what 
is the minimum level of amplitude in the $B$ modes which is detectable in
presence of foregrounds. 
Although preliminary investigations concerning the minimum detectable level of 
primordial tensors 
exist \citep{tucci_etal_2005}, no satisfactory answer has been given yet, essentially because 
of the foreground uncertainties \citep{baccigalupi_2003}. Algorithms aiming at the 
reduction of the foreground contamination in CMB observations
belong to the category of the component separation techniques. Those are designed 
to use multi-frequency information to separate emissions observed in the same 
frequency bands but produced by different physical processes. If robust 
prior knowledges are available about the signal to recover, the maximum entropy method 
\citep{hobson_etal_1998,stolyarov_etal_2002} or Wiener filtering 
\citep{tegmark_efstathiou_1996,bouchet_etal_1999} may be implemented. 
On the other hand, in the case of CMB polarization, the polarized 
foregrounds are likely to be greatly uncertain even at the time when the 
analysis of the future CMB data will be already on-going, and thus alternative 
approaches, not relying on such priors, may be required. The class of ``blind" 
component separation techniques exploits the statistical independence 
of the sky signals to be separated, a natural expectation for the CMB 
and Galactic emissions. \\
Among the techniques in this category, the Independent Component 
Analysis (ICA) \citep{amari_chichocki_1998,hyvarinen_1999} has been considered 
in several works concerning component separation. It was first 
exploited as a neural network, i.e. capable of 
self-adjusting on time varying data streams \citep{baccigalupi_etal_2000}, 
and then developed in a form of a numerically fast algorithm and a code, {\fastica},
capable of operating on a data set as nominally anticipated from the Planck 
experiment 
\citep{maino_etal_2002}. The latter approach has been successfully tested on the 
real data from COBE/DMR \citep{maino_etal_2003}, recovering the 
main CMB results of that experiments, e.g., with respect to the amplitude 
and power spectrum of the cosmological perturbations on large 
scales, and also on the data of the  BEAST experiment \citep{donzelli_etal_2005}. A flexible
version of the ICA algorithm, capable to exploit 
available priors, has been proposed \citep{delabrouille_etal_2003}. 
Recently, the {\fastica} algorithm was applied to simulated Planck data 
in polarization on all sky \citep{baccigalupi_etal_2004}. 

In the forthcoming years, the detection of the $B$ modes will be 
attempted by balloon and ground based experiments, 
targeting sky regions where the foreground emission in total intensity 
is known to be low. One of those regions is centered on the position 
$(40^{\circ},-45^{\circ})$ in right ascension and declination. 
This region has been observed by BOOMERanG 2K \citep{montroy_etal_2005}, 
and is the target of the EBEx \citep{oxley_etal_2004}, QUAD 
\citep{bowden_etal_2004} and QUIET experiments\footnote{For a list of the 
operating and planned CMB polarization experiments see {\tt lambda.gsfc.nasa.gov}.}.
However, a low amplitude of the foreground emission in total 
intensity does not ensure that on a level of the anticipated cosmological $B$-mode 
signal the polarized foregrounds can {\it a priori} be considered irrelevant.
Therefore in this work we test the 
{\fastica} performance in the reconstruction of the CMB polarized emission on that 
region of the sky, focusing on the recovery of $B$ modes. Our work represents 
an exploratory study, not specialized to describe any particular operating or 
planned experiment, nor to quantify what is the minimum level of $B$ modes detectable 
in presence of foregrounds. Our aim is rather to determine
if the blind component separation techniques have the capability to recover the $B$ modes 
of the CMB, which are on the level as predicted in the current 
best fit cosmological models and are observed on a limited patch of the sky
in the presence of a substantial foreground contamination as estimated on the basis of
the current models of the Galactic emission.
In section \ref{sm} we describe how the background and 
foreground signals are simulated, while in section \ref{ppps} we describe how 
the angular power of the CMB polarization anisotropies on a limited 
portion of the sky is computed. 
In section \ref{cs} we present a parallel version of the {\fastica} algorithm 
({\lightica}) operating on polarization data and apply it to the simulated 
sky signal realization to evaluate 
biases and stability of the $B$ mode reconstruction against variations 
in the background and noise realizations, noise and foreground fluctuation 
amplitude and extension of the sky area considered. 
In section \ref{mtpta} we discuss the implications of our results 
for what concerns the recovery of the primordial tensor to scalar ratio.
Finally, in section \ref{d} we discuss and summarize our results. 

\begin{figure}
\begin{center}
\includegraphics[width=8cm]{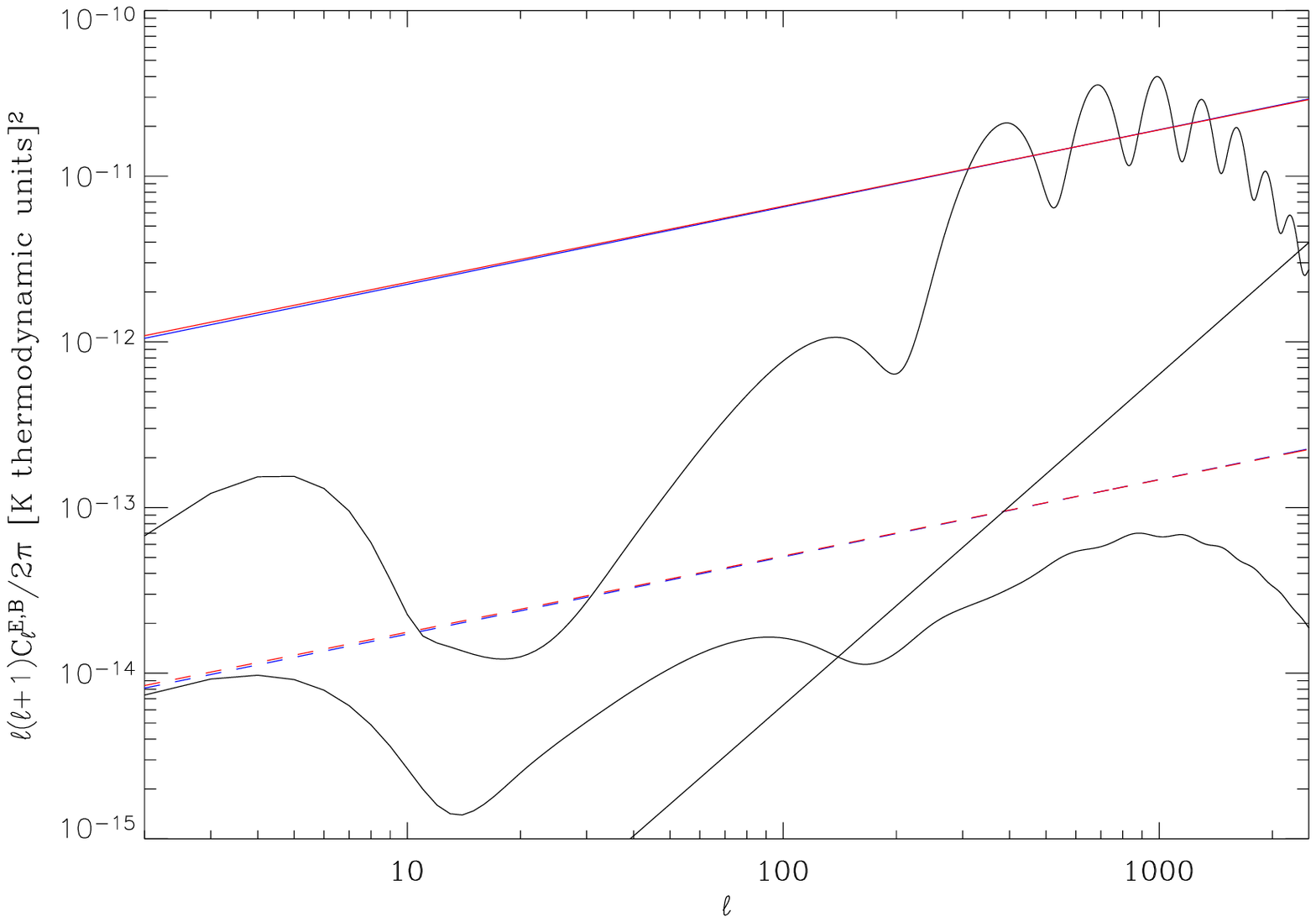}
\includegraphics[width=8cm]{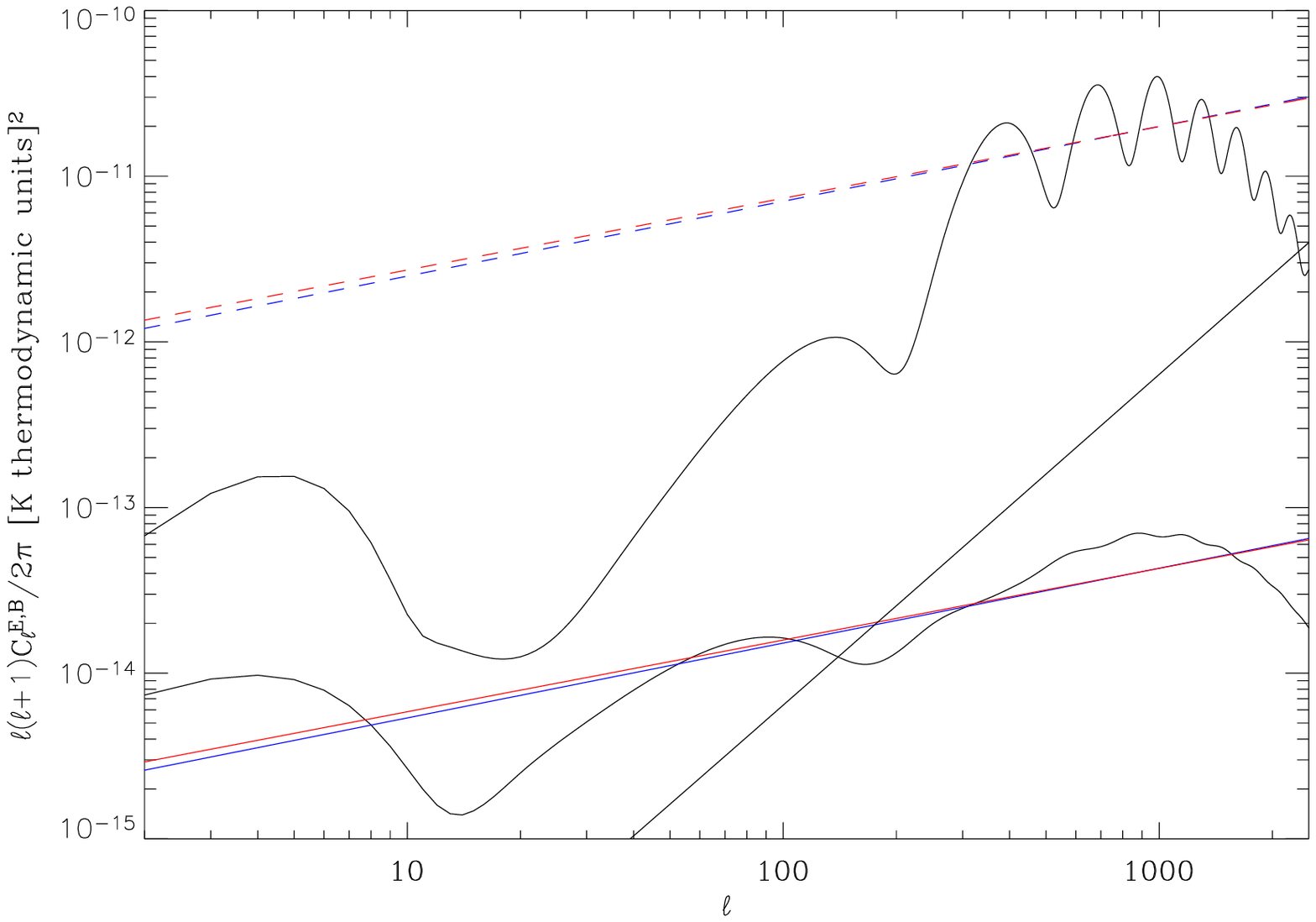}
\end{center}
\caption{Power spectra of the different polarized sky signals relevant to the 
microwave observations.
The almost flat straight lines represent the foreground contamination 
obtained by cutting out the Galactic plane up to $|b|=50^{\circ}$, and fitting
those with a power law; the steep straight lines raising as $l^{2}$ represent
the instrumental noise 
assumed in this work. The left and right panels show the predictions for 
the 40, 90 GHz and 150, 350 GHz  frequency bands, respectively. 
The solid lines represent a lower frequency, 
while the dashed ones a higher one. These signals are plotted against the full sky 
CMB power spectra of anisotropies in thermodynamical units for $E$ and $B$ as 
assumed in this work (oscillating solid curves).}
\label{b50contamination}
\end{figure}

\section{Simulated maps}
\label{sm}

The CMB emission is simulated accordingly to the cosmological 
concordance model \citep{spergel_etal_2003}. The Hubble constant 
is $H_{0}$=72 km/s/Mpc, the overall geometry is flat, with 
a critical density made of baryons (4.4\%),
Cold Dark Matter (CDM) (22.6\%), and the cosmological constant (73\%).
The radiation component consists of photons and three massless neutrino 
species. The optical depth to the last scattering surface is fixed at $\tau =0.11$. 
The perturbations are Gaussian, with a primordial power spectrum 
characterized by a spectral index of scalar perturbations $n_s=0.96$.
Unless otherwise specified, the primordial gravity wave contribution is set to
10\% of the scalar perturbation amplitude, with a spectral index fixed accordingly 
to the single field inflationary model, $n_{t}=-n_{s}/6.8$. When explicitly 
specified, we also consider the case in which no gravity waves are present.
We include the contribution due to lensing in the power spectrum, which is 
responsible for substantial part of the power in the $B$ modes of the CMB 
polarization anisotropies. The statistics of this component is non-Gaussian 
because of the correlation of different scales induced by lensing itself 
\citep{smith_etal_2004}, and that could be exploited to separate it from 
the primordial $B$-mode signal \citep{seljak_hirata_2004}. On the other hand, 
as we show in the next Section, on a limited patch of the sky
considered here, the angular power spectrum for $B$ is substantially affected by 
the leakage coming from the $E$ modes and which thus obey the Gaussian statistics of
the primordial fluctuations.
Given this and the fact that, in the following we consider only 2-point statistics,
we assume that the entire $B$-mode signal is Gaussian.
The power spectrum of CMB anisotropies is calculated using 
{\sc cmbfast} \citep{seljak_zaldarriaga_1996} while the sky realizations 
have been obtained using the HEALPix package 
\footnote{http://www.eso.org/science/healpix/}, assuming the Gaussian 
statistics. In antenna units, 
the CMB fluctuations at a frequency $\nu$ are obtained by the 
thermodynamical one by multiplying by the factor 
$x^{2}\exp{(x)}/[\exp{(x)}-1]^{2}$, where $x=h\nu /kT_{CMB}$, $h$ and $k$ 
are the Planck and the Boltzmann constant, respectively, while 
$T_{CMB}=2.726$ K is the CMB thermodynamical units. 

The polarized synchrotron emission has been simulated 
accordingly to two distinct recipes. Both of them 
derived a template for the polarization angle $\theta$ by exploiting 
the observations in the radio band: these measures indicate a 
rather high fluctuation level interpreted as the effect of 
the small scale structure of the Galactic magnetic field 
\citep{uyaniker_etal_1999,duncan_etal_1999}, scaling as 
$C_{l}^{\theta}\sim l^{-2}$ on degree and sub-degree 
angular scale, up to the arcminute \citep{tucci_etal_2002}, 
consistently also with recent observations at medium Galactic latitudes 
\citep{carretti_etal_2005}.
It is worth noting here that WMAP three year analysis shows an evidence for 
a shallower slope in the polarization angle pattern, at least on 
large angular scales and intermediate Galactic latitudes \citep{page_etal_2006}. 
The template for the polarization angle was obtained by adopting the form 
above for $C_{l}^{\theta}$, and assuming Gaussian distribution. 
The distinction between the two recipes is in the model assumed for the 
polarized intensity. In \citet{baccigalupi_etal_2001}, it was 
derived directly from the observations in the radio band 
including the existing data on large angular scale 
\citep{brouw_spoelstra_1976}. On the other hand, \citet{giardino_etal_2002} 
exploited the all sky template of synchrotron in total 
intensity at 408 MHz \citep{haslam_etal_1982}, 
assuming a theoretically synchrotron polarization fraction of
about 75\%; since the latter template has a resolution of about 
one degree or less, they extrapolated the power to the smaller scales 
by exploiting the total intensity observations in the radio band 
\citep{uyaniker_etal_1999,duncan_etal_1999}. The recipe adopted 
by \citet{giardino_etal_2002} yields a 
stronger signal, and is what we take in this work as the 
polarized synchrotron template. In antenna units, 
the frequency scaling of the synchrotron is related to the 
energy distribution of electrons, exhibiting a steep power 
law as $\nu^{-3}$, according to the observations of WMAP 
at intermediate and high Galactic latitude \citep{bennett_etal_2003}
and as adopted in this paper.

The polarized emission from the diffuse thermal dust has been 
detected for the first time in the Archeops data \citep{benoit_etal_2004}, 
indicating a 5\% polarization fraction with respect to the total intensity 
emission, which is very well known at 100$\mu$m and can be extrapolated at 
microwave frequencies fitting for the emissivity and temperature 
of two thermal components \citep{finkbeiner_etal_1999}. In this work, 
we adopt the model 8 of \citet{finkbeiner_etal_1999}, 
where dust emissivity and temperatures do not vary across the sky. 
The dust polarization fraction reported by WMAP three years is also 
consistent with a few percent. 
The pattern of the polarization angle is much more uncertain, and due 
to the magnetized dust grains which get locally aligned along the 
Galactic magnetic field \citep{prunet_etal_1998,jones_etal_1995}. 
Since the geometry and composition of the dust grains is 
still very uncertain, the simplest assumption is that the 
Galactic magnetic field is 100\% efficient in imprinting the 
polarization angle pattern to the synchrotron and dust emission 
\citep{baccigalupi_2003}. 

Although the knowledge of the polarized foregrounds summarized 
above may be exploited to build their all sky models, 
the resulting templates, though useful are still affected by a substantial 
uncertainty, and hence need to be utilized with a caution. 
For synchrotron, the uncertainties are mainly due to a poor resolution
of the available total intensity template \citep{haslam_etal_1982}, which corresponds
to an angular scale of a degree or larger, and, which in addition is polluted 
by HII regions at low Galactic latitudes \citep{baccigalupi_etal_2001}. 
Moreover, the polarized signal is observed in the radio band only, and at 
low Galactic latitudes, where a substantial Faraday depolarization may 
significantly affect the true synchrotron pattern.
Indeed, as we already mentioned, WMAP three years reports evidence of a shallower 
slope in the power spectrum of polarized fluctuations of diffuse foregrounds, 
although that claim is limited essentially by sensitivity to large angular 
scales and intermediate Galactic latitudes \citep{page_etal_2006}. 
The dust model is much better known in total intensity, but the polarization 
fraction has again been measured at low Galactic latitudes only, and on large angular 
scales. Moreover the dust polarization angle distribution suffers the same uncertainty 
as in the synchrotron case, as in the modeling the two are commonly assumed 
to be identical. 

\begin{figure}
\begin{center}
\includegraphics[width=8cm]{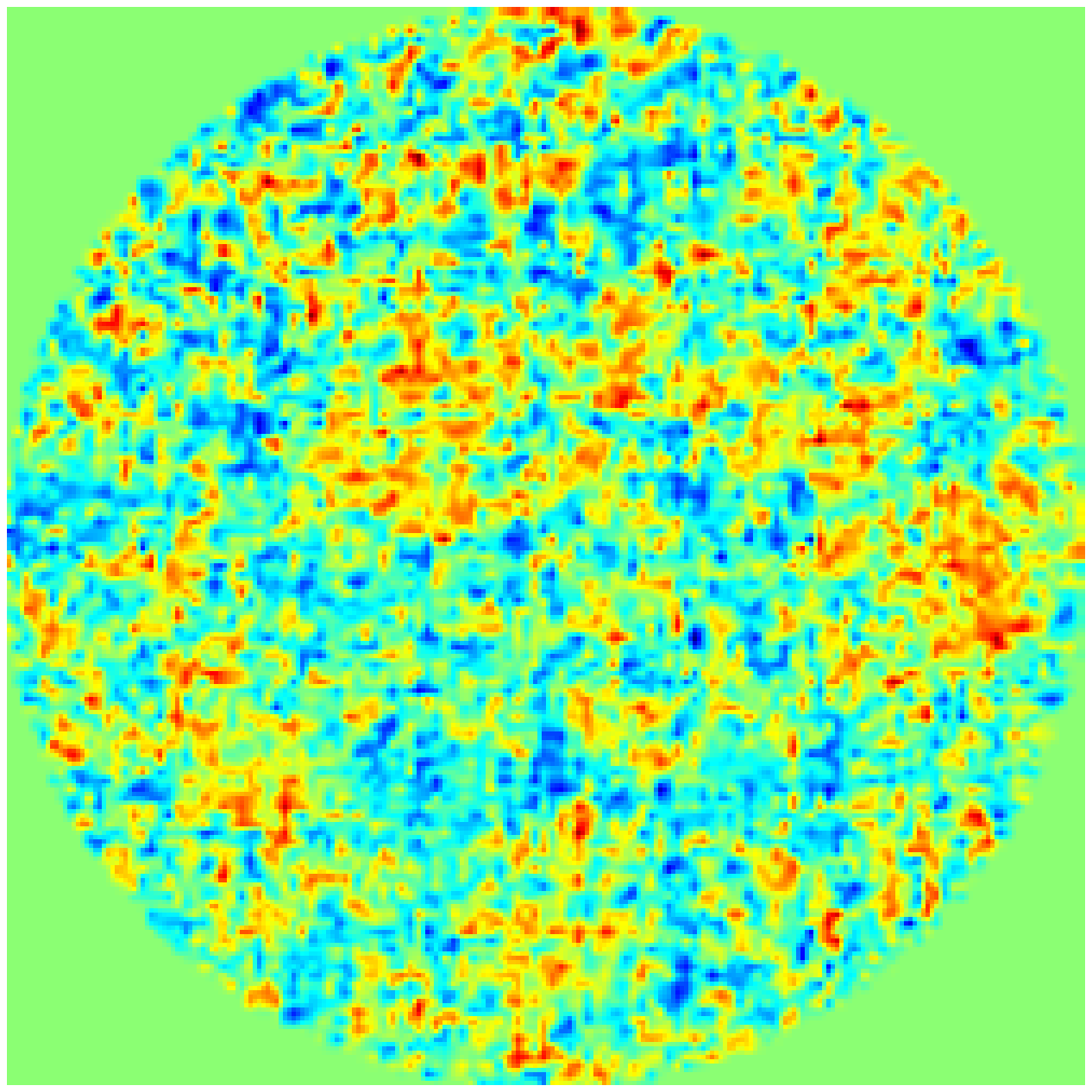}
\includegraphics[width=8cm]{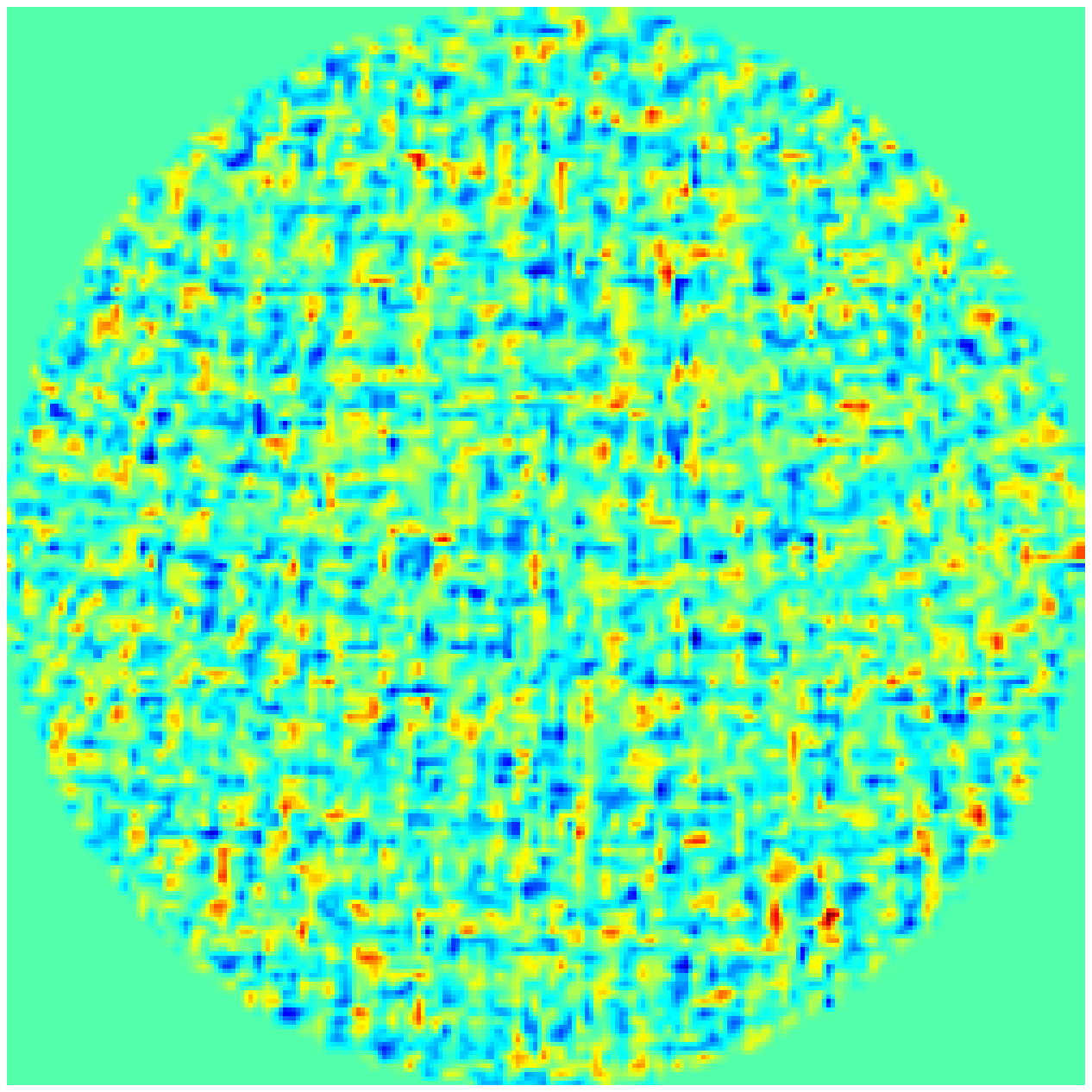}
\end{center}
\caption{The total (CMB plus foregrounds) $Q$ Stokes parameter emission in the 
sky area 
considered in this work, at 40 (left) and 90 GHz (right). 
At 90 GHz the signal appears dominated by the CMB signal, 
while the synchrotron contamination is evident at 40 GHz.}
\label{Q_syn}
\end{figure}

\begin{figure}
\begin{center}
\includegraphics[width=8cm]{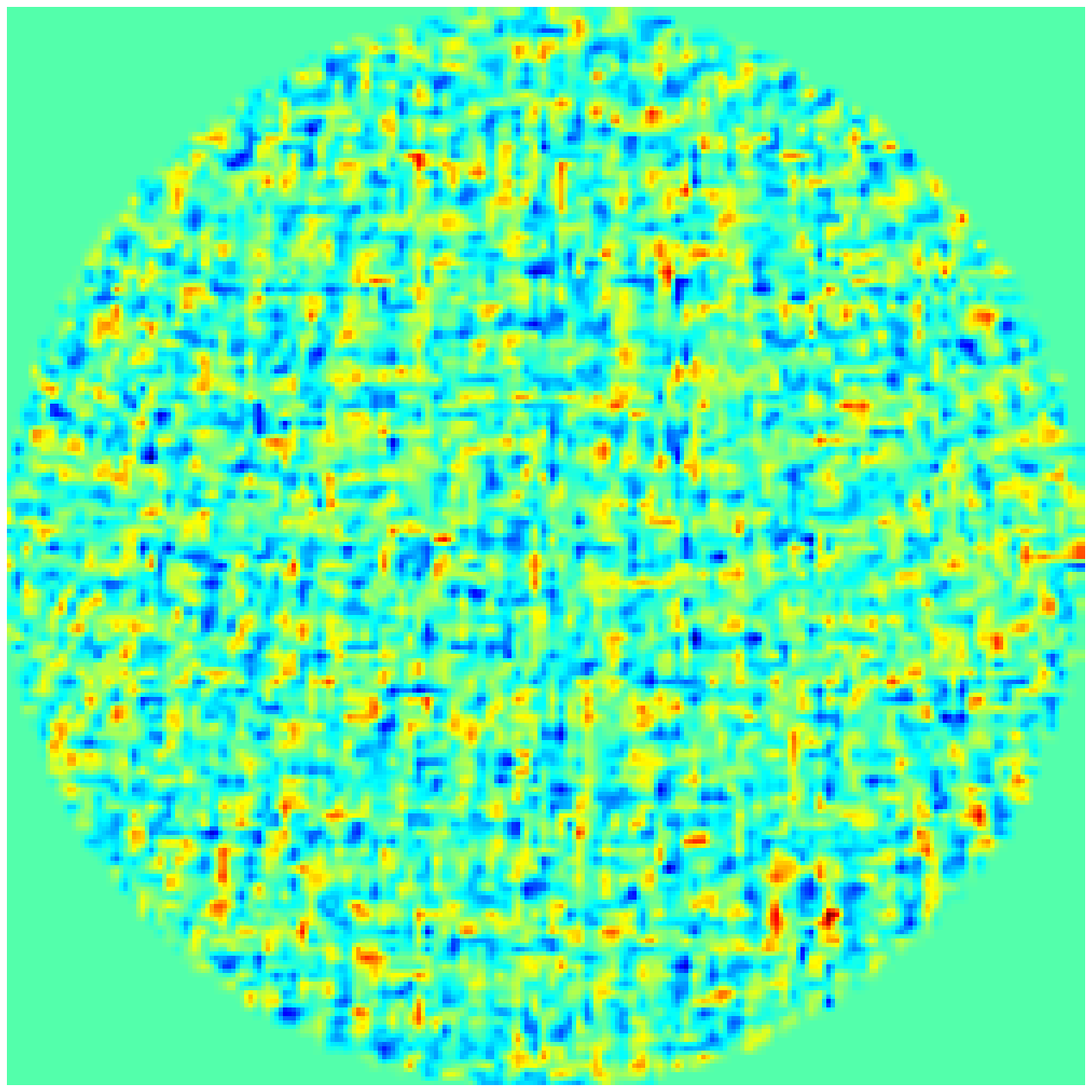}
\includegraphics[width=8cm]{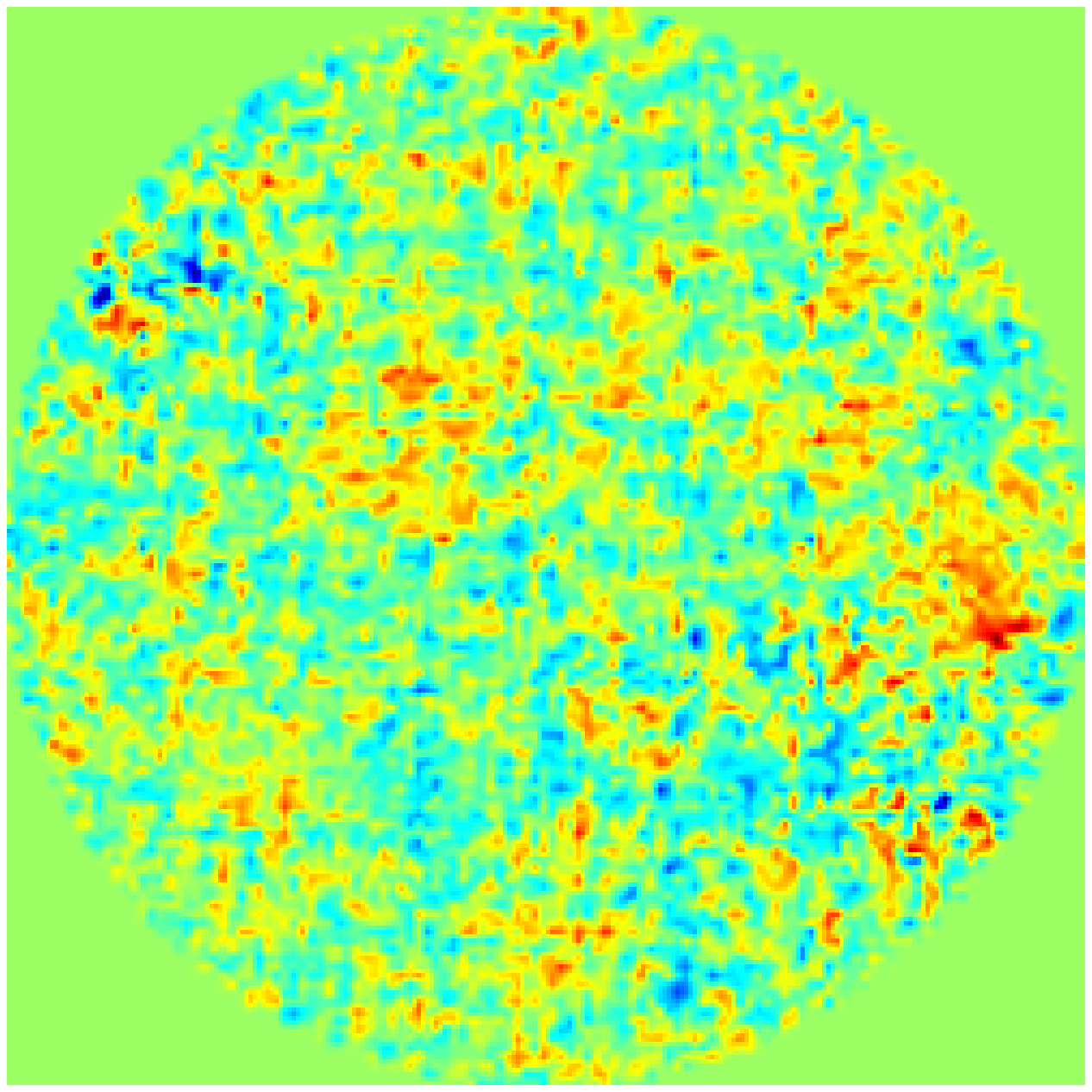}
\end{center}
\caption{The total $Q$ Stokes parameter emission in the sky area 
considered in this work, at 150 (left) and 350 GHz (right). 
At 150 GHz the signal appears dominated by the CMB emission, 
while the dust emission dominates at 350 GHz.}
\label{Q_dust}
\end{figure}

Despite of these missing pieces, this represents the present state of the art in 
the simulation of the polarized Galactic foreground emission,
at least on scales and Galactic latitudes still hidden to WMAP. 
The forecasted contamination is particularly challenging for the $B$ mode CMB 
measurements; 
indeed, as this signal, arises only from the primordial gravitational waves
and the weak lensing effect of the $E$ polarization 
\citep{zaldarriaga_seljak_1997,zaldarriaga_seljak_1998}, it is 
about one order of magnitude smaller than the $E$ component. 
On the other hand, the Galactic foregrounds are expected 
to have almost the same power in the two modes \citep{zaldarriaga_2001}, 
as WMAP three year results confirmed remarkably. \\
In figure \ref{b50contamination} we show the contamination to the all sky 
CMB $E$ and $B$ spectra from the foreground emission corresponding to the 
synchrotron and dust diffuse Galactic signal after cutting out the Galactic 
plane up to $|b|=50^{\circ}$, 
roughly corresponding to the latitudes considered in this work, as we explain 
below. The foreground power has been evaluated by fitting the actual 
sky signal with a power law, $C_{l}=\alpha l^{\beta}$. As it is evident, 
the models of the foreground emission indicate that the contamination to the 
$B$ modes of the CMB is relevant in all cases. The lines raising as $l^{2}$ 
represent the levels of instrumental noise which we consider in this 
work, as explained in detail in section \ref{cs}. \\ 
In this work we study the performance of the ICA technique on diffuse signals 
in polarization, and we do not consider the effect of extra-Galactic point sources, 
although this point is certainly crucial in realistic conditions; indeed the
current models \citep{tucci_etal_2004} suggest that the residual power from
unresolved point sources could be comparable to the level of noise we treat
here, at least at the lowest frequency we considered, and although the results
we show next are quite stable against the noise amplitude, 
this point warrants a further investigation in forthcoming works. We come back
to this point in the conclusions. \\
We shall consider a circular sky patch with a radius 
$\theta_C=10^{\circ}$ and $20^{\circ}$, corresponding to about $0.76\%$ and 
$3.04\%$ of the entire sky, respectively. The center in Galactic coordinates is 
at $l=260^{\circ}$, 
$b=-62^{\circ}$, within the region considered by different experiments 
\citep{montroy_etal_2005,bowden_etal_2004,oxley_etal_2004}. We take two 
frequency combinations, 40, 90 GHz, and 150, 350 GHz, where the dominant 
foreground emission is represented by the synchrotron and the thermal dust, 
respectively. The sky emission at the various frequencies, corresponding to 
the $Q$ Stokes parameter, is shown in figures \ref{Q_syn} \&\  \ref{Q_dust}. 
At 90 and 150 GHz the CMB signal appears relatively free of foreground 
contamination, while at 40 and 350 GHz the foregrounds dominate. In Section 
\ref{cs} we describe more quantitatively the foreground CMB contamination by 
means of the angular power spectrum, defined formally in the next Section when 
a limited part of the sky is considered. 

\section{Polarization pseudo power spectra}
\label{ppps}

In this work we apply the ICA component separation technique on
a portion of the sky. We will quantify the quality of the reconstruction
with help of the angular (pseudo-)power spectra, which are relevant and straightforwardly
calculable for the limited sky observations as considered here.
In a computation of the polarized, $E$ and $B$ (pseudo-)power spectra
on a finite portion of the sky a transfer of power between the $E$  and $B$ 
modes occurs \citep[see][ and references therein]{chon_etal_2004}.
Since the $B$ modes are sub-dominant, the 
leakage of  the $E$-mode power alters their spectrum more substantially
and consequently needs to be explicitly considered in the presented analysis.
In this paper we will denote the pseudo power spectra 
of $E$ and $B$ as $\tilde{C}_l^E$ and $\tilde{C}_l^B$, 
respectively, while the symbols without a tilde will correspond to their full-sky versions.
Hereafter we compute the power spectra 
using a recipe adopted from \citet{hansen_etal_2002}. Consequently,
we introduce a window function, $G(\theta,\phi )$, \citep{gabor_1946} which is applied to the data
prior to a computation of the spherical harmonic transforms on a portion of the sphere and
a calculation of the pseudo-power spectra.
The leakage between the polarization modes may be written as 
\begin{equation}
\label{pseudoe}
\tilde{C}_l^E\;=\;\sum_{l^{\prime}} C_{l^{\prime}}^E 
K_2(l,l^{\prime})+\sum_{l^{\prime}} C_{l^{\prime}}^B K_{-2}(l,l^{\prime})\ ,
\end{equation}
\begin{equation}
\label{pseudob}
\tilde{C}_l^B\;=\;\sum_{l^{\prime}} C_{l^{\prime}}^B K_2(l,l^{\prime})+\sum_{l^{\prime}} 
C_{l^{\prime}}^E K_{-2}(l,l^{\prime})\ ,
\end{equation}
where $C_{l}^E$ and $C_{l}^B$ are the polarization full-sky power spectra 
\citep{zaldarriaga_seljak_1997}, while the kernels $K_2(l,l^{\prime})$ 
and $K_{-2}(l,l^{\prime})$ depend on the form and the size of the cut, 
described by a generic function $G(\theta,\phi )$ which is zero in the 
sky regions which are not considered. 
The explicit expressions for the kernels are:
\begin{equation}
\label{kernels}
K_{\pm2}(l,l^{\prime})\;=\;\sum_{l^{\prime\prime}} 
g^2_{l^{\prime\prime}}\frac{(2l^{\prime}+1)(2l^{\prime\prime}+1)}
{32\pi^2}W^2(l,l^{\prime},l^{\prime\prime})(1\pm(-1)^{l+l^{\prime}+
l^{\prime\prime}})\ .
\end{equation}
Here $g_l$ are found by the inverse Legendre transform of the Gabor window 
$G(\theta,\phi)$ and the Wigner symbols $W$ are defined as:
\begin{equation}
W(l,l^{\prime},l^{\prime\prime})=
\left(
\begin{array}{ccc}
l&l^{\prime}&l^{\prime\prime} \\
-2&2&0
\end{array}
\right)\ .
\end{equation}

We exploited these formulae for circular cut sky area of different size with 
top hat shape: 
\begin{equation}
G(\theta)=
\left\{
\begin{array}{l l}
{\displaystyle 1,} & {\displaystyle \theta\leq\theta_C}\ ,\\
{\displaystyle 0,} & {\displaystyle \theta > \theta_C}\ .
\end{array}
\right.
\end{equation}
As one can see from equations (\ref{pseudoe}) and (\ref{pseudob}), the sky cut 
mixes the polarization $E$ and $B$ modes, as quantified by the
$K_{-2}(l,l^{\prime})$ kernel. Obviously, the mixing gets reduced as the 
size of the window is increased. Since the cosmological fluctuations are 
dominated by the scalar contribution in the cosmological concordance model 
\citep{spergel_etal_2003}, even if the 
diagonal of the kernel $K_2(l,l^{\prime})$ is one order of magnitude larger 
than the diagonal of $K_{-2}(l,l^{\prime})$, we expect the $E$ mode to 
contaminate substantially the $B$ signal even for large regions of the 
sky, while on the other hand $\tilde{C}_l^E \simeq C_l^E$. 

\begin{figure}
\begin{center}
\includegraphics[width=8cm]{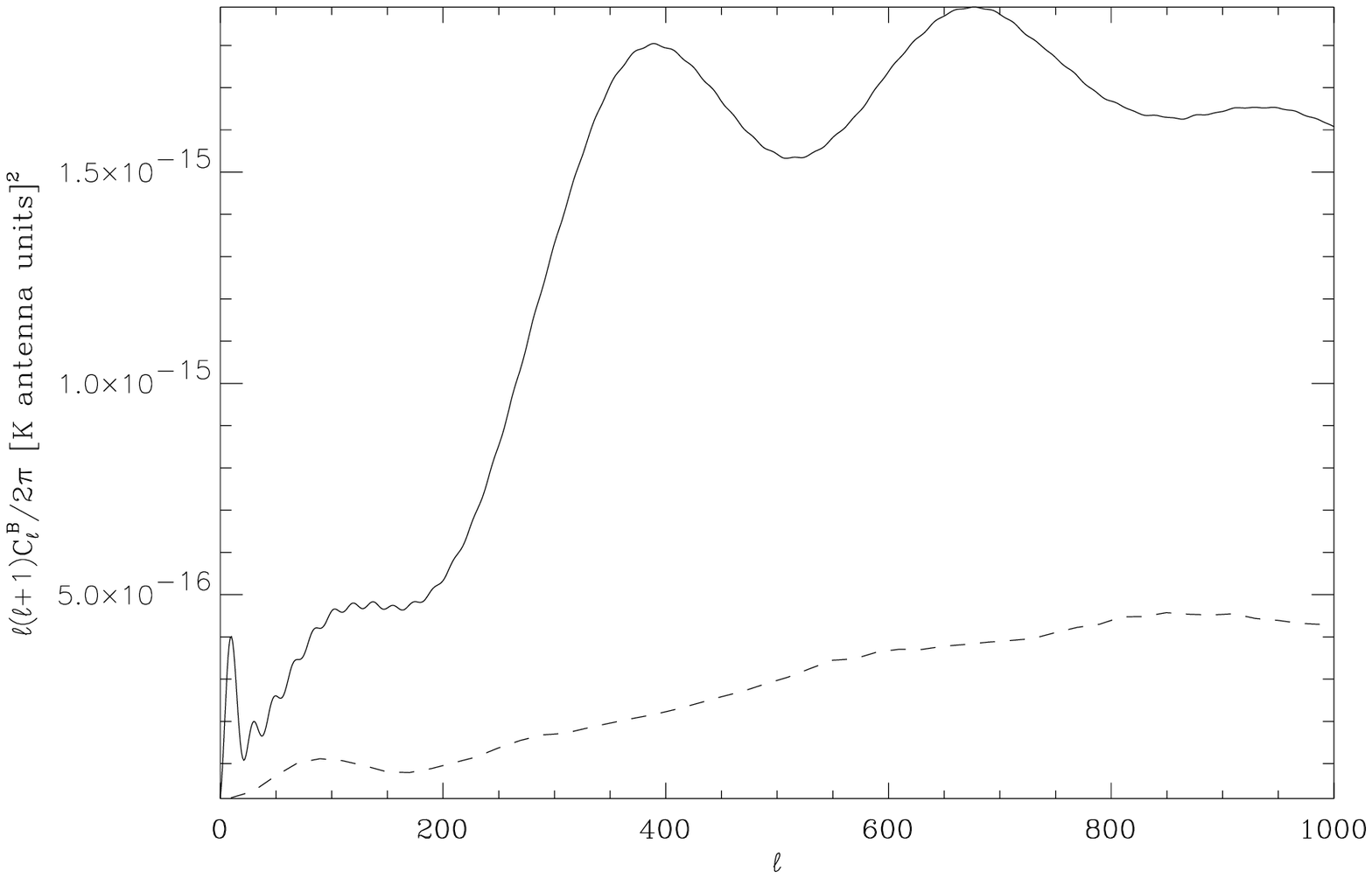}
\includegraphics[width=8cm]{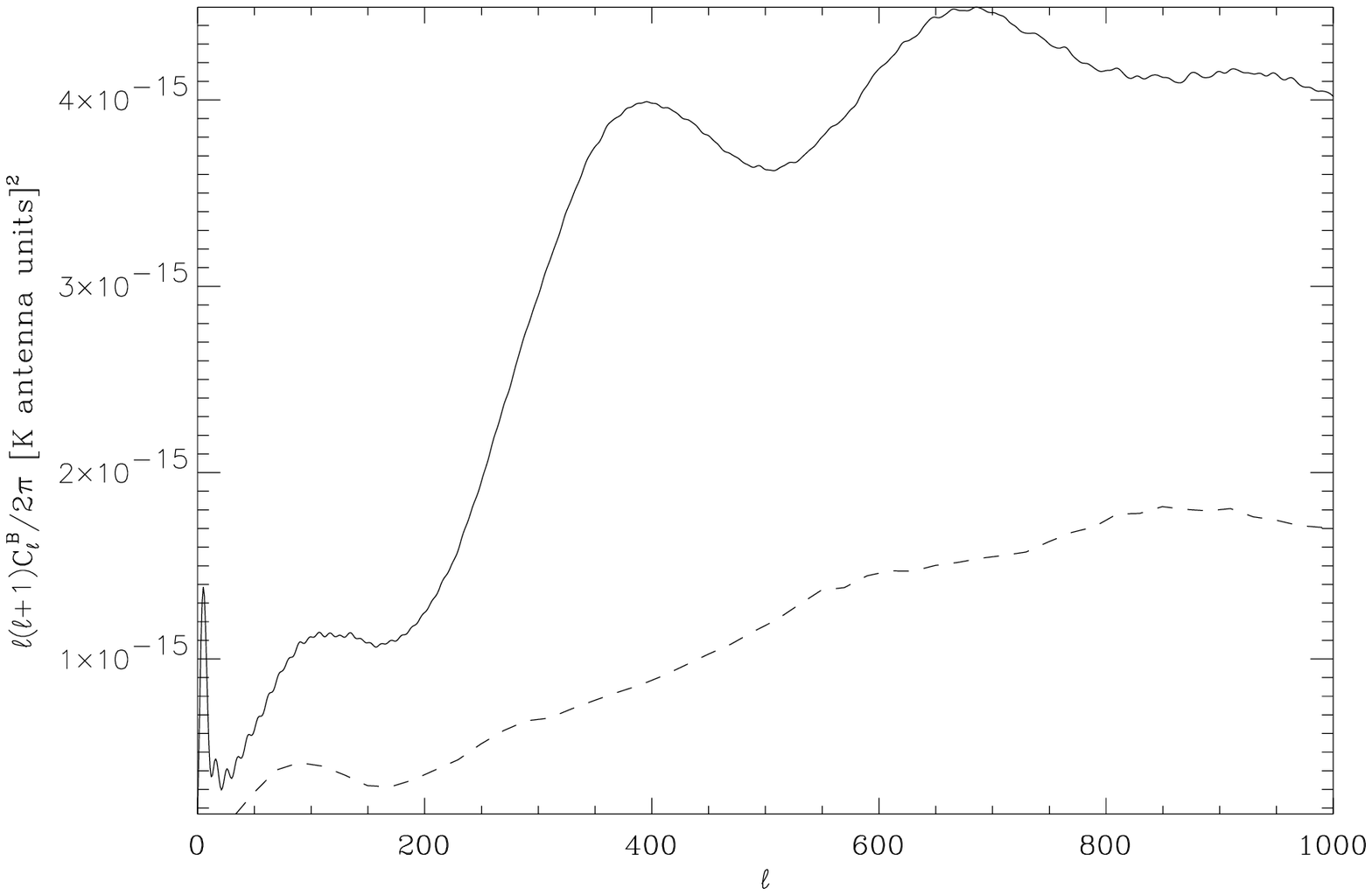}
\end{center}
\caption{Pseudo power spectra for the $B$ modes, in the case of a circular top hat 
cut of $\theta_{C}=10^{\circ}$ (left panel, solid line) and 
$\theta_{C}=20^{\circ}$ (right panel, solid line). The dashed lines in both 
panels represent the full sky $C_l^B$ normalized to the patch area fractions. 
The contamination due to the $E$ mode is evident.}
\label{bmodes1545}
\end{figure}

In Fig.\ref{bmodes1545} we show the pseudo power spectrum of the $B$ mode, 
$\tilde{C}_l^B$, as defined in (\ref{pseudob}) for a top hat window with 
$\theta_C=10^{\circ}$ and $20^{\circ}$. In the latter case, the leakage 
from the $E$ modes is slightly weaker. (Hereafter, we limit our analysis to a range of
$l$-modes $\le1000$ in order to speed up  the calculation of the pseudo-$C_{l}$s.)
For comparison the dashed lines show the 
full sky $B$ mode power spectra (with power normalized on the patch). As we 
see, the shapes of the two spectra are substantially different and the $E$ 
mode contamination is relevant for the pseudo-$C_l^B$.

\section{Component separation}
\label{cs}

In this Section we present a first application of the {\lightica} 
code, based on {\fastica}, optimized for parallel runs and operating 
on polarization data. We first describe the basic features of the algorithm, 
then show its performance on our simulated data. We do not report here 
all the details of the {\fastica} technique, as they may be found in earlier 
works, concerning total intensity \citep{maino_etal_2002} 
and polarization \citep{baccigalupi_etal_2004}. As described above, we consider 
here two frequency combinations, 40 and 90 GHz, 150 and 350 GHz, where 
the CMB is contaminated by synchrotron and dust diffuse polarized emission, 
respectively. The CMB is added to the foregrounds at the different 
frequencies in $Q$ and $U$, expressed in Kelvin antenna units. We adopt 
a spatial resolution of about $3.5'$ arcminutes, corresponding to $n_{side}=1024$ 
resolution parameter in the HEALPix pixelization scheme. Before processing, all the 
maps have been smoothed with a Gaussian, circular beam with full width 
half maximum (FWHM) equal to 10 arcminutes; moreover, a Gaussian and 
uniformly distributed noise is added, with amplitude specified below. 
As we stress in the following, the high resolution is a crucial requirement for a 
successful application of the ICA techniques, as they do not rely on any other priors 
than the statistical independence of the signal to reconstruct, and need 
a large number of realizations to converge to the solution. 
The spatial resolution considered here is close to the values anticipated for the 
forthcoming experiments
\citep{montroy_etal_2005,bowden_etal_2004,oxley_etal_2004}. \\
In all the cases we show, a single separation run (not including 
the construction of the simulated sky) takes about 5 seconds on a 
workstation with 1.5 Gb RAM and 2.4 GHz Pentium IV processor.

\subsection{{\lightica:} an {\sc MPI} parallel implementation of the  {\fastica} component separation in polarization}
\label{tlcfcsip}

The core of the ICA technique is based on a maximization of 
an approximation of the neg-entropy, which measures the distance 
of a mixture of signals from a Gaussian distribution 
\citep{amari_chichocki_1998,hyvarinen_1999}. The hypothesis are 
that the mixture contains at most one Gaussian 
component and that all of them obey different probability distributions 
and frequency scalings; if this is verified, it is possible to demonstrate that 
asymptotically in the number of realizations (pixels in our case), 
the local maxima of the neg-entropy correspond to the components 
present into the mixture. The excellent performance obtained so 
far on simulated CMB data is due mainly to two features: the good 
detail in the maps, reaching the arcminute scale, as well as 
the high level of independence between the background CMB and the 
foreground emissions \citep{maino_etal_2002,baccigalupi_etal_2004}.
As reported in the latter reference the {\fastica} technique can be, and has been,
applied  to sky maps of $Q$ and $U$ Stokes parameters, and
which subsequently can be used to derive constraints on the $E$ and $B$ power spectra.\\ 
Given those promising results, and aiming at the application of the method
to real polarization data, it is important to develop a framework for
evaluating the errors of the separation as well as for studying its performance 
for a broad set of values of basic parameters characterizing
any data set. The relevant parameters in this case include instrumental noise, CMB realizations, foreground 
properties, area covered by a given experiment and others. Along with pure 
theoretical consideration, these two issues may be addressed with the help of suitable Monte Carlo 
simulations. Due to the remarkable speed of a single ICA component separation run, 
this technique is well suited for straightforward parallelization with only a
memory management and I/O issues calling for more elaborate solutions. 
In the {\lightica} code, the main ICA routine, operating on arrays of
numbers, is called by each processor processing its own version of the sky (or
part thereof). 
The code consists in a main driver performing the following steps. A number 
of CMB realizations is pre-computed and a set of external parameters determines
the case under study, namely set of frequencies, foreground fluctuation 
amplitude, noise amplitude, etc. For each CPU, first, the driver generates random numbers 
representing random noise in the data. Then it calls a separate routine co-adding the 
different sky components as defined by the external run parameters.
The sky components are represented in the HEALPix format and the co-addition
is done separately for the $Q$ and $U$ Stokes parameters.
Let ${\bf x}^{Q}$ and ${\bf x}^{U}$ 
be the multi-frequency data, where ${\bf x}$ is labeled by two indices, numbering 
frequencies (rows) and pixels (columns). The algorithm assumes that the 
components scale rigidly 
with frequency, which means that each of them can be represented by a 
product of two functions each depending either on a frequency or a direction (i.e., a pixel). 
We define a spatial pattern for them,
which we denote with either ${\bf s}^{Q}$ or ${\bf s}^{U}$ and
express the inputs ${\bf x}^{Q,U}$ as
\begin{equation}
\label{xQU}
{\bf x}^{Q,U}={\bf A}^{Q,U}\, {\bf s}^{Q,U}+{\bf n}^{Q,U}\ ,
\end{equation}
Here the matrices ${\bf A}^{Q,U}$ scale the spatial patterns of
${\bf s}^{Q,U}$ to the input frequencies; the instrumental noises 
${\bf n}^{Q,U}$ have same dimensions as ${\bf x}$. Note that 
equation (\ref{xQU}) implies that all the frequencies has the same 
spatial resolution, which is not the case in real observations and is 
one of the limitations of the ICA technique in pixel space, 
effectively forcing the analysis to be performed at the lowest resolution 
of a given experiment. \\
The constructed skies are given as inputs to the 
main ICA core routine performing the separation. 
The maximization of the neg-entropy computes two separation matrices, 
${\bf W}^{Q}$ and ${\bf W}^{U}$, and produces a copy of the independent components 
present in the data:
\begin{equation}
\label{yQU}
{\bf y}^{Q}={\bf W}^{Q}{\bf x}^{Q}\ \ ,\ \ {\bf y}^{U}={\bf W}^{U}{\bf x}^{U}\ .
\end{equation}
More details on the way the separation matrix for {\fastica} is
estimated are given in previous works 
\citep{maino_etal_2002,baccigalupi_etal_2004}, together 
with a recipe how to recover the frequency scaling of the signals 
${\bf s}^{Q,U}$. The resulting ${\bf y}^{Q}$ and ${\bf y}^{U}$ can be combined 
together to get the $E$ and $B$ modes for each reconstructed component. 
Note that the noise correlation matrix can be taken into account in 
the separation process; this is done simply by subtracting the noise correlation 
matrix from the one of the total signal before entering into the core of the 
algorithm \citep{hyvarinen_1999}; for an uniform and 
Gaussian distributed noise, its correlation matrix is null except on the 
diagonal, containing the noise variances at the frequencies considered. 
Of course even in the case of a perfect separation, the derived
outputs will be noisy as implied by equations (\ref{xQU},\ref{yQU}).
Consequently, whenever the separation matrices, ${\bf W}^{Q, U}$, are 
well-constrained by data, the dominant contribution to the noise level is 
adequately approximated by,
\begin{equation}
\label{nyQU}
{\bf n}_{y}^{Q,U}={\bf W}^{Q,U}{\bf n}^{Q,U}\ .
\end{equation}
This means that, if the noises for different channels are uncorrelated 
and Gaussian, and denoted as $\sigma_{\nu_{j}}$ the input noise root 
mean square (${\it rms}$) 
at frequency $\nu_{j}$, the noise ${\it rms}$ on the $i$-th output is\begin{equation}
\label{sigmayQU}
\sigma_{y_{i}}^{Q,U}=
\sqrt{\sum_{j}|W_{ij}^{Q,U}|^{2}|\sigma_{\nu_{j}}^{Q,U}|^{2}}\ .
\end{equation}
For full sky signals, the noise contamination to the angular 
power spectrum is 
$C_{l,\, n_{i}}^{Q,U}=4\pi (\sigma_{y_{i}}^{Q,U})^{2}/N$, where
$N$ is the number of pixels. The Gaussianity and uniformity assumptions
make it easy to calculate the noise level on $E$ and $B$ modes,
since they contribute at the same level: 
$C_{l,\, n_{i}}^{E}=C_{l,\, n_{i}}^{B}=
(C_{l,\, n_{i}}^{Q}+C_{l,\, n_{i}}^{U})/2$. 
The latter quantities represent the average noise power, which can be
simply subtracted from the output power spectra by virtue of the
lack of correlation between noise and signal. The remaining uncertainty 
comes from noise realization, which at $1\sigma$ and on the whole sky is: 
$\Delta C_{l,\, n_{(i)}}^{E,B}=\sqrt{2/(2l+1)}C_{l,\, n_{i}}^{E,B}$. \\
The final step consists in the output of the results. Those may be 
in the form of maps or power spectra, computed simply with the HEALPix 
routines. The code also outputs the separation matrix and its inverse 
as computed by each processor. \\
The overall structure of the {\lightica} is rather flexible; in particular, 
the header dealing with different variables may be easily 
changed and specialized for studying a particular degree of freedom.
It is also quite extensible and, for example, the {\sc openMP} HEALPix routines can
be easily incorporated if a further
speed-up of the power spectrum computation is desired.\\
In the following we present the first applications of {\lightica}. 
We choose and analyze a suitable reference simulated dataset, 
and then we study the stability of the 
results against variation of some among the most relevant degrees of freedom of 
the simulated dataset. In each case, the separation quality is quantified by the 
ICA induced bias and additional uncertainty of the recovered CMB spectra as 
evaluated in each Monte Carlo series. 

\subsection{$B$ modes reconstruction and error estimation}
\label{bmraee}

\begin{figure}
\begin{center}
\includegraphics[width=11cm]{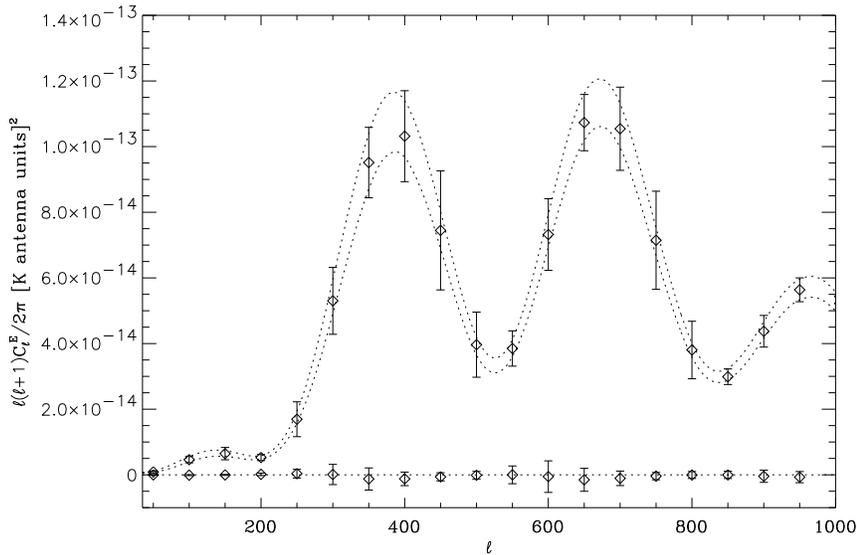}
\end{center}
\caption{Pseudo-power spectra of the reconstructed $\tilde{C}_{l}^{E}$ modes of the CMB in the 40, 90 GHz
frequency combination, in the $S/N=2$ case. The region between
the dotted lines is the theoretical CMB signal $\pm\sigma$ cosmic 
and noise variance at 40 GHz on the sky area considered.
At the bottom we show the average and standard 
deviation of the residuals on each realization.}
\label{oute_noi05_syn}
\end{figure}

\begin{figure}
\begin{center}
\includegraphics[width=11cm]{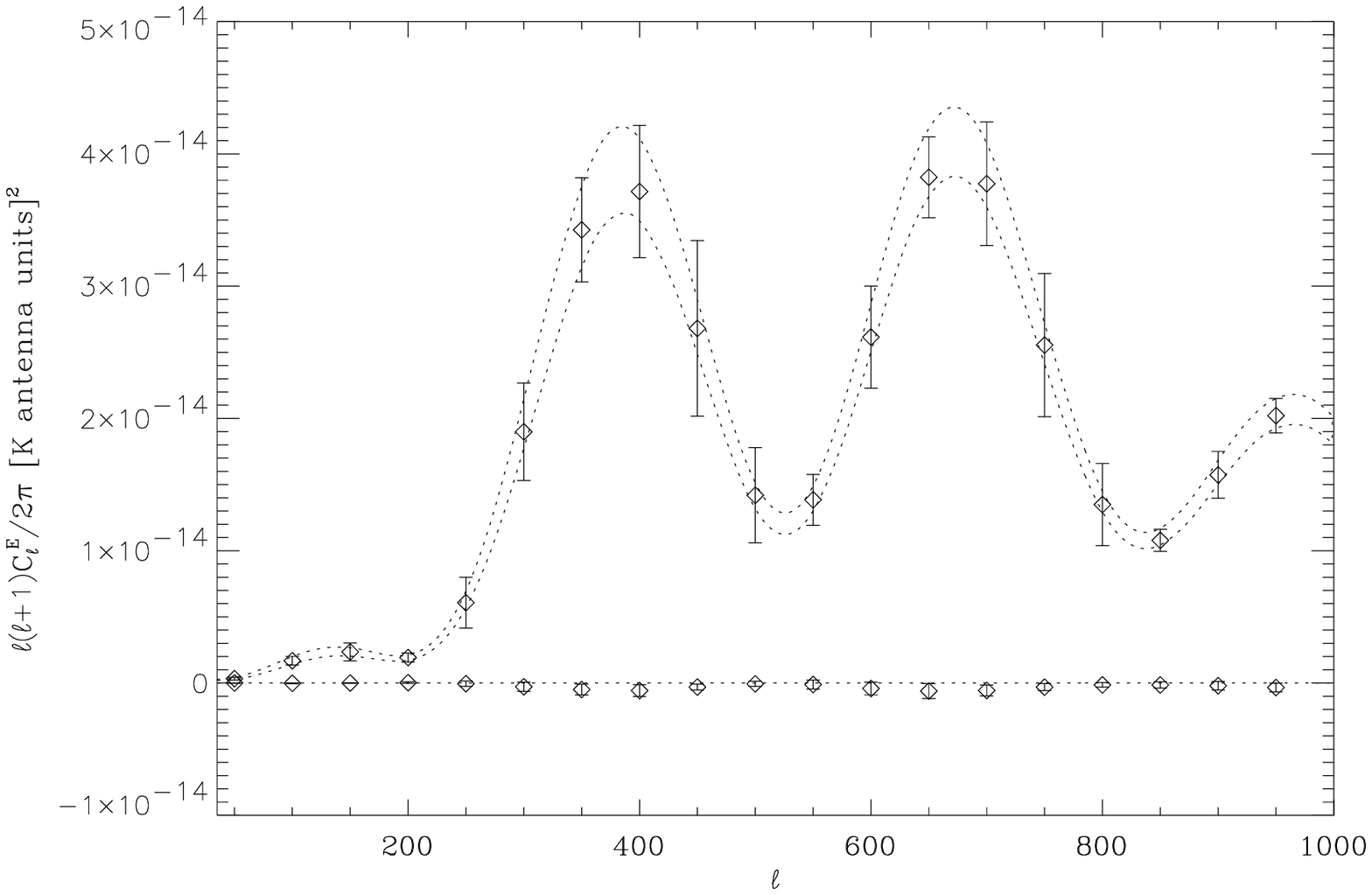}
\end{center}
\caption{Pseudo-power spectra of the reconstructed  $\tilde{C}_{l}^{E}$ modes of the CMB in the 150, 350 GHz
frequency combination, in the $S/N=2$ case. The region between
the dotted lines is the theoretical CMB signal $\pm\sigma$ cosmic 
and noise variance at 150 GHz on the sky area considered.
At the bottom we show the average and standard 
deviation of the residuals on each realization.}
\label{oute_noi05_dust}
\end{figure}

\begin{figure}
\begin{center}
\includegraphics[width=11cm]{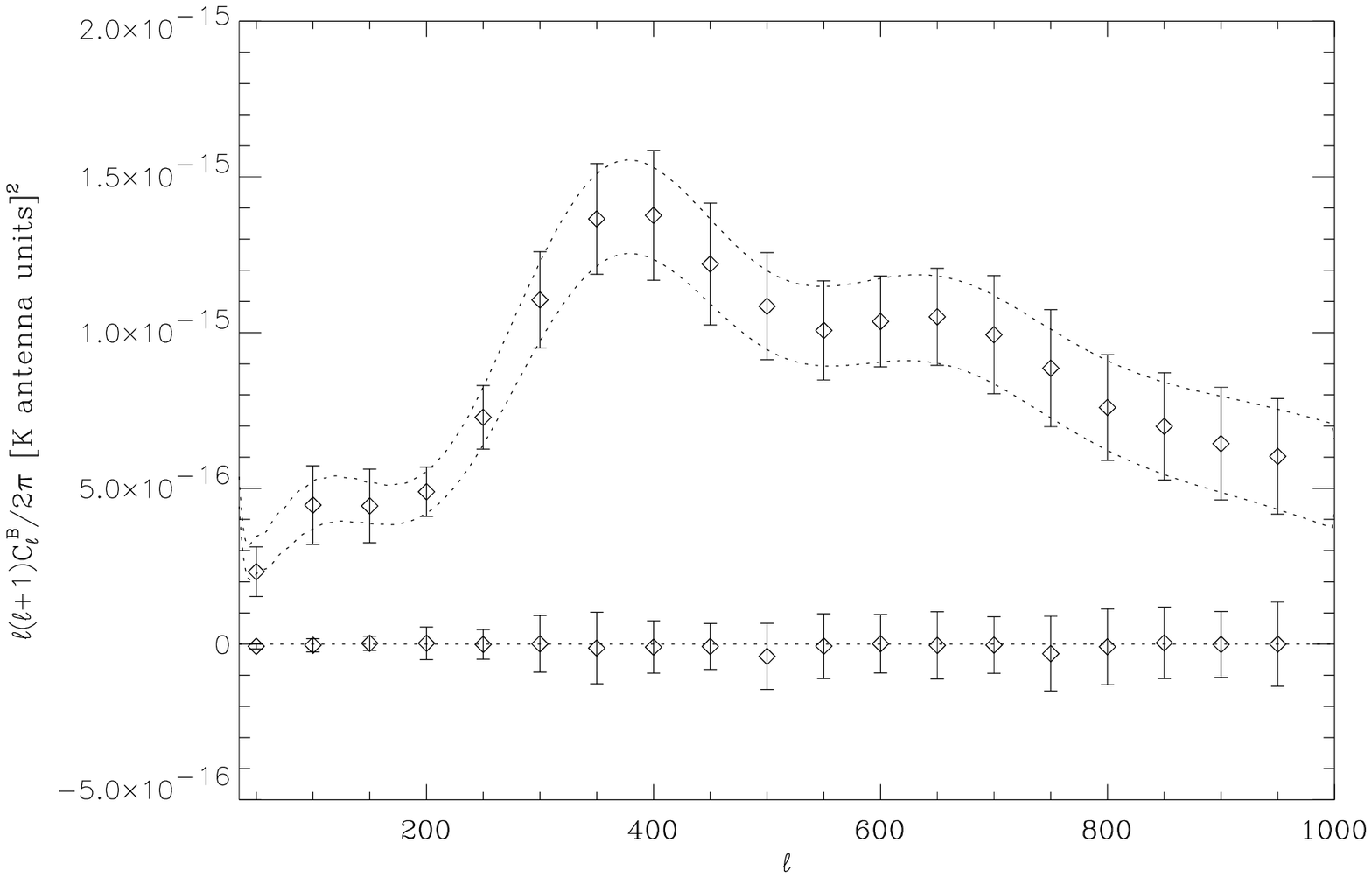}
\end{center}
\caption{Pseudo-power spectra of the reconstructed  $\tilde{C}_{l}^{B}$ modes of the CMB in the 40, 90 GHz
frequency combination, in the $S/N=2$ case. The region between
the dotted lines is the theoretical CMB signal $\pm\sigma$ cosmic 
and noise variance at 40 GHz on the sky area considered.
At the bottom we show the average and standard 
deviation of the residuals on each realization.}
\label{outb_noi05_syn}
\end{figure}

\begin{figure}
\begin{center}
\includegraphics[width=11cm]{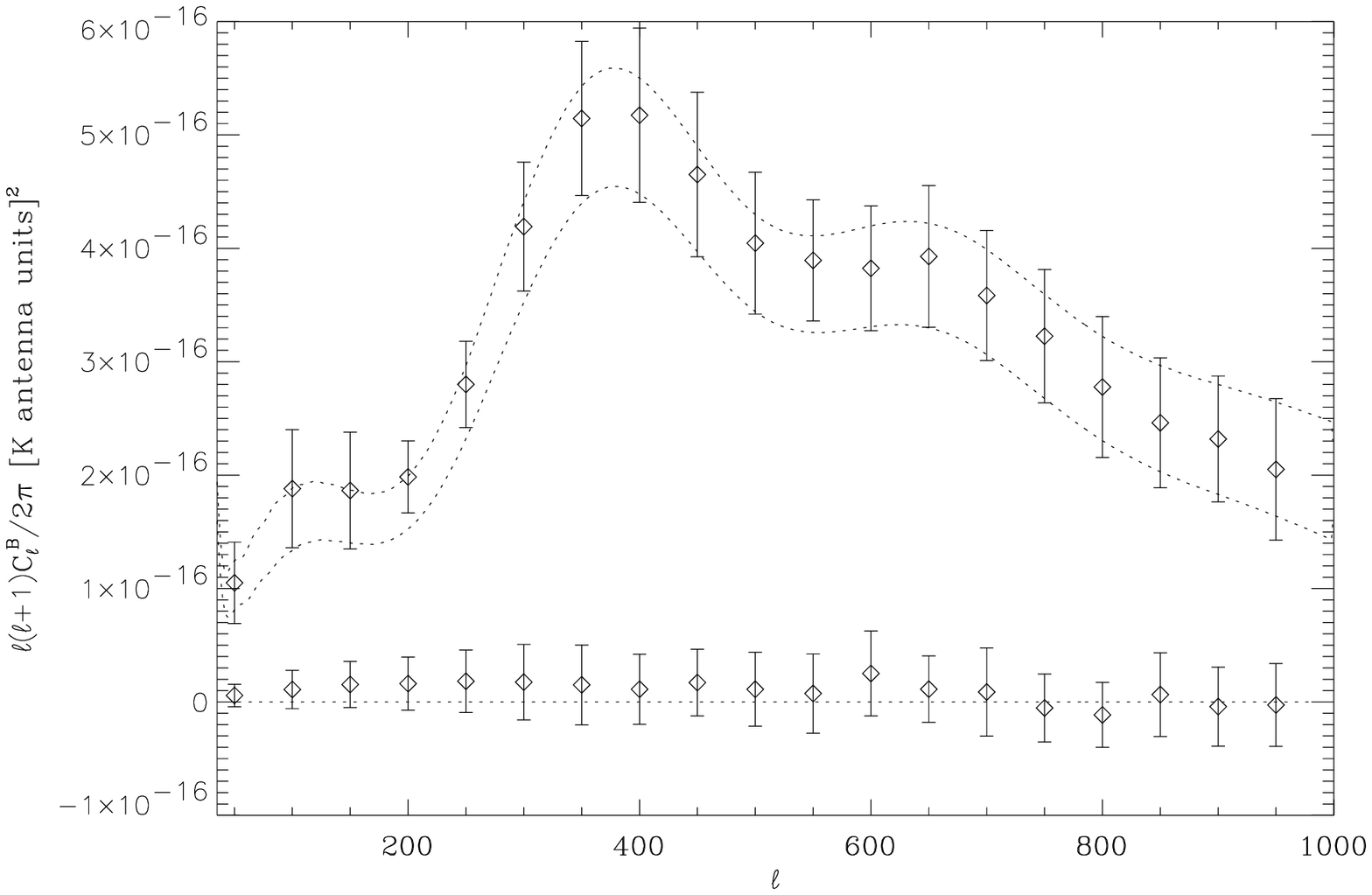}
\end{center}
\caption{Pseudo-power spectra of the reconstructed $\tilde{C}_{l}^{B}$ modes of the CMB in the 150, 350 GHz
frequency combination, in the $S/N=2$ case. The region between
the dotted lines is the theoretical CMB signal $\pm\sigma$ cosmic 
and noise variance at 150 GHz on the sky area considered.
At the bottom we show the average and standard 
deviation of the residuals on each realization.}
\label{outb_noi05_dust}
\end{figure}

\begin{figure}
\begin{center}
\includegraphics[width=8cm]{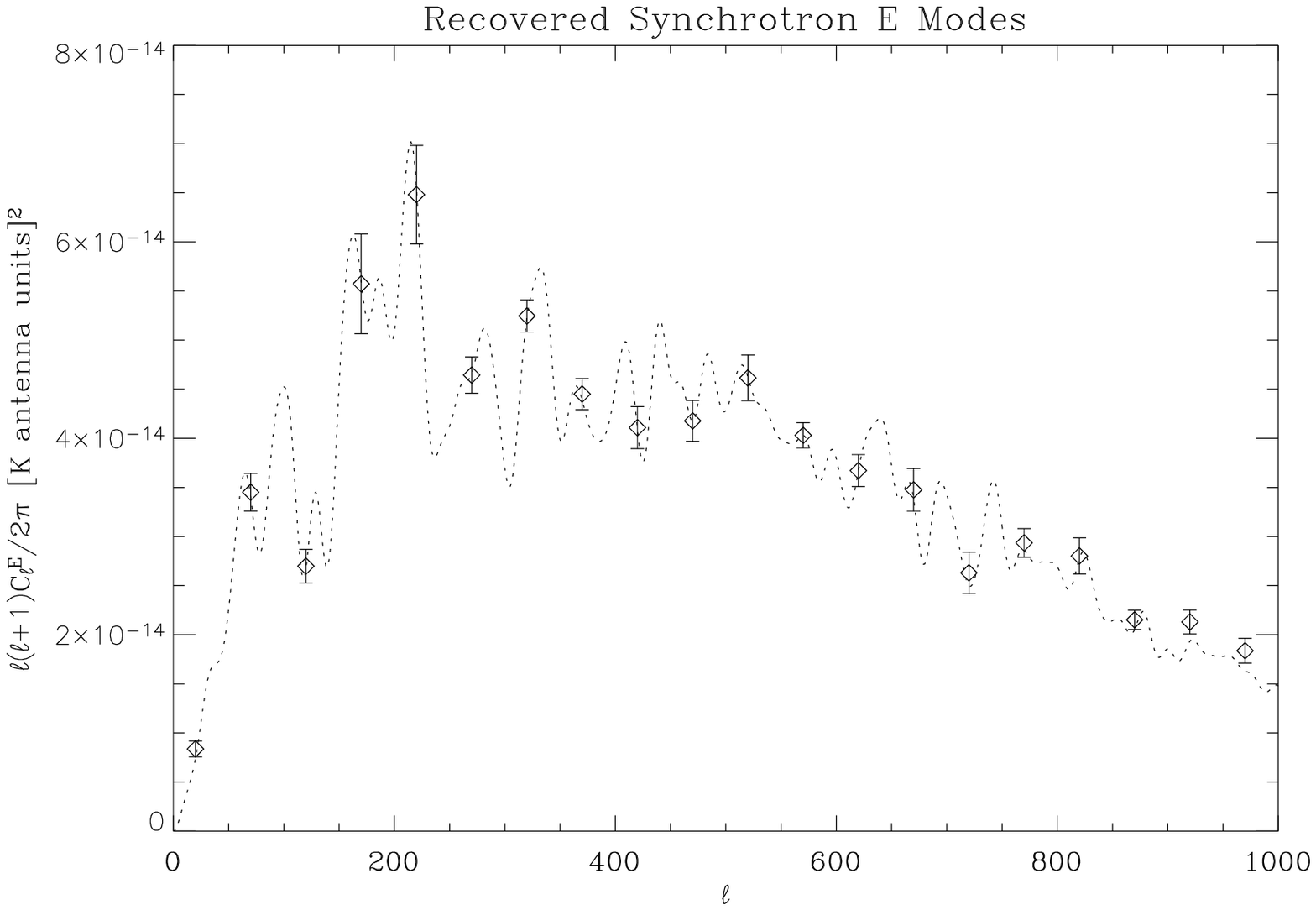}
\includegraphics[width=8cm]{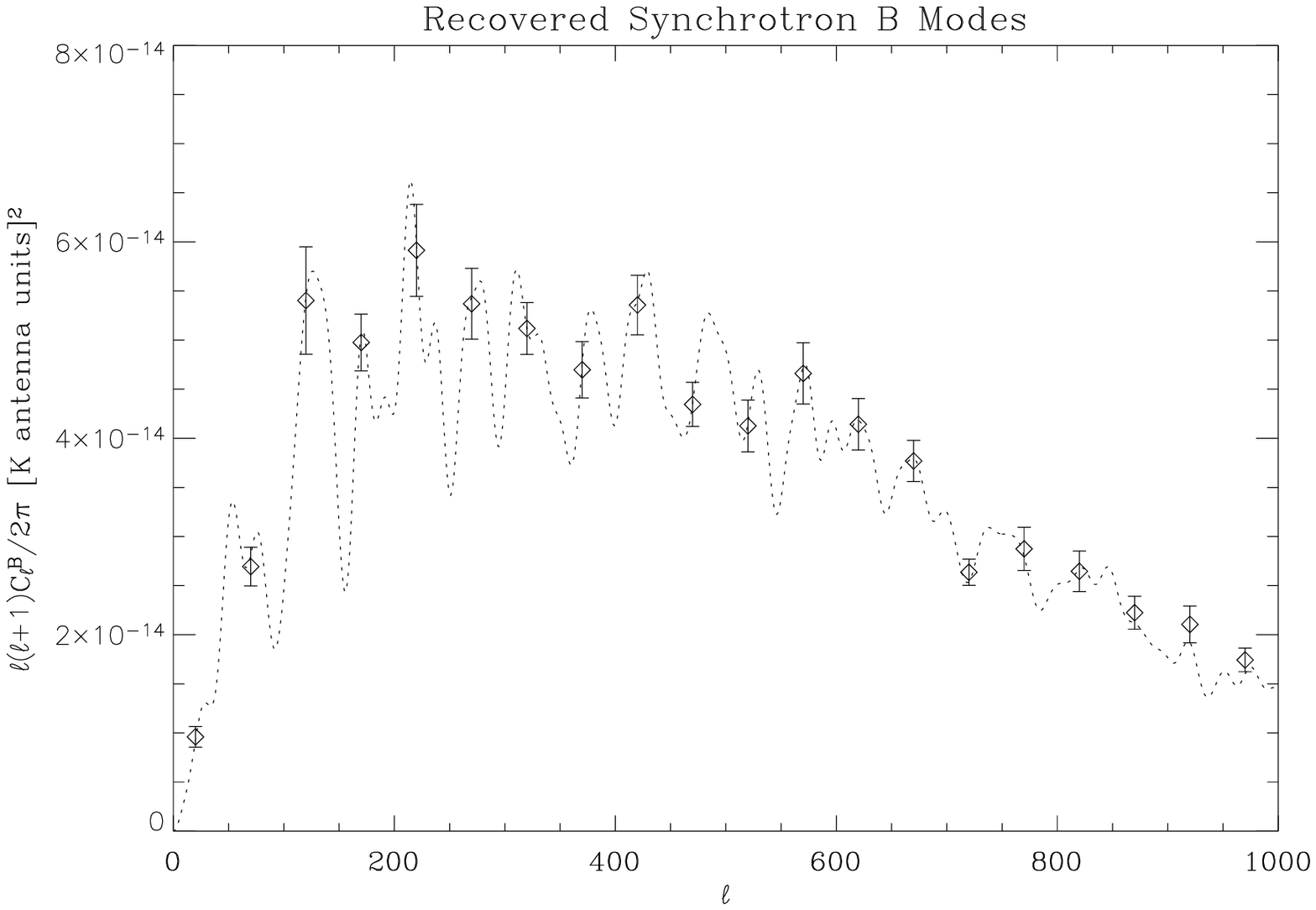}
\end{center}
\caption{Pseudo-power spectra of the reconstructed $\tilde{C}_{l}^{E}$ modes 
(left) and $\tilde{C}_{l}^{B}$ (right) of synchrotron at 40 GHz
in the $S/N=2$ case. Dotted lines are the theoretical spectra.}
\label{rec_fore}
\end{figure}

The sky signals in the patch considered are processed by the {\lightica} 
code, and the outputs, in $E$ and $B$, are shown in figures 
\ref{oute_noi05_syn}, \ref{oute_noi05_dust}, and \ref{outb_noi05_syn}, 
\ref{outb_noi05_dust}, respectively. 
Those are plotted at 40 and 
150 GHz in antenna units, as the code outputs are at the lowest 
frequency by default. 
In each panel, the two dotted curves correspond to the theoretical 
pseudo-$C_{l}^{E,B}$ of the CMB signal, $\pm 1\sigma$ where $\sigma$ 
represents the cosmic variance on our patch of the sky: that is specified by 
a fraction $f_{sky}$ and binned over $\Delta l=50$ multipoles \citep{tegmark_1997}, 
and is given by 
\begin{equation}\label{cosmvar}
\Delta{\tilde{C}_{l}^{E,B}}=
\sqrt{\frac{2}{(2l+1)\Delta l f_{sky}}}(\tilde{C}_{l}^{E,B}+
{\tilde{C}_{n,\, l}^{E,B}})\ ,
\label{cosmvar2}
\end{equation}
where $\tilde{C}_{n,\, l}^{E,B}$ are the contribution of the noise. 
We assume a Gaussian and uniformly distributed noise over the analyzed region, 
with {\it rms} equal to a half of that of the CMB Stokes parameter Q or U 
on a single pixel, 
at each frequency. The noise amplitude is not related to any particular experiment, 
and was chosen as a starting point for the analysis performed in the next 
sub-section, where the noise amplitude is varied. Note that, in the case of
full sky coverage, 
the chosen noise amplitude is shown in figure \ref{b50contamination}. \\
The symbols in the figures represent the signal recovered by the {\lightica} separation 
process averaged over the 100 MC simulations of the CMB and noise, while the
error bars show, a $1\sigma$ uncertainty derived from the simulations.
Thus, they represent the error in the separation process, 
given the foreground templates assumed in this work.
At the bottom of each figure, we also plot the average and standard 
deviation of the residuals, obtained 
by subtracting the input from the output pseudo-power spectra for each realization.
The averages provide a measure of biases of the reconstruction 
on each realization, while the error bars estimate the extra dispersion introduced
due to the separation process. \\
The first feature to be noted is that the separation is clearly successful, 
for $E$ and $B$ as well. 
Note that the B-mode pseudo-power spectra are generally comparable or lower than the foreground 
and noise contamination, as we show explicitly in the next sub-section. 
As we stressed above, the ICA technique looks for the independent components 
into the data, assuming rigid and different frequency scaling and a different 
statistics for all of them, with no other prior; the fact that this procedure 
is able to extract with such a precision a signal which is comparable or lower 
than the foreground contamination in presence of noise is remarkable. 
Once again, the observed performance is made possible by the large number of pixels 
in the map, as well as the high level of statistical independence between 
background and foreground emission. These two facts 
bring the algorithm close to an ideal environment, ensuring the convergence very 
close to the correct answer, with a precision represented by the errors shown 
in the figures. \\
A second, most interesting aspect to be noted is that we detect the error due 
to the separation process; that is clearly visible in all the figures as the 
excess in the error bars with respect to what predicted by cosmic variance 
and noise. The error from component separation is comparable or smaller, 
to the sample variance of the simulated templates. 
The error of the separation
is either due to the randomness of the noise realizations on one hand and the fact
that, for a single 
realization, background and foreground may not be completely independent.
The latter factor can be a source of the extra randomness in the ICA performance 
thus contributing to the total error.\\
Although in this work we are mainly interested in the extraction 
of the CMB $B$ modes from the data, it is interesting also to look 
at the foreground recovery. In figure 
\ref{rec_fore} we plot the reconstructed pseudo power spectra of the 
separated synchrotron compared with the original ones, reported with 
dotted lines. In this case the {\lightica} is able to properly reconstruct 
the polarized signals of the foreground with good precision. On the other 
hand, the dust reconstruction fails, as it comes out heavily contaminated 
by the CMB, and with wrong normalization. This manifests that 
the separation with dust is more problematic, as it may be also noted by looking 
at figure \ref{outb_noi05_dust}, which shows excess power in the recovered spectra 
and residuals with respect to the input ones, and which is mostly concentrated at low multipoles 
where the dust spectrum is highest, see also figure \ref{foreground_rms_dust}. 
This occurrence should not be interpreted in terms of the different pattern 
of the foreground emission for dust and synchrotron, but in terms of the 
relative weight of it with respect to the background emission, as already 
noticed in earlier works \citep{maino_etal_2002,baccigalupi_etal_2004}. 
Due to the difference in the frequency scalings in the bands considered, 
in the 40, 90 GHz case the foreground and background signals are 
closer in amplitude with respect to the higher frequency combination; 
thus, at 150 GHz the CMB dominates over the dust while at 350 GHz 
the CMB emission is negligible. 
Indeed, this bias disappears if the foreground amplitude is raised by a 
factor of a few as we see next, and consequently the dust template 
can be better reconstructed.\\
Finally, note that our Monte Carlo analysis does not include 
varying the foreground template, a factor should be accounted for as well 
in order to quantify the error in the separation process in a comprehensive way. 
However, the modest knowledge of the foreground emissions as it is
available at the present does not allow 
to estimate their statistics to a level high enough to vary their template 
in the Monte Carlo.


\subsection{Varying noise, foreground amplitudes and analyzed area}
\label{vnafaaa}

We perform here a first study of a dependence of the results on some of
the key simulation parameters. Specifically, we vary 
noise amplitude, foreground fluctuation amplitude, and extension 
of the sky area considered. We explore the corresponding parameter space by
moving along the multiple ``directions" within its volume and use the results to
set constraints on the applicability 
of the {\lightica} approach given the assumed foreground pattern, however,
still in systematic free cases. \\
In order to quantify the error introduced by the algorithm with respect to 
the one due to the cosmic variance and noise, and focusing on the 
$B$ mode reconstruction, we introduce the quantities 

\begin{equation}
\label{ratios}
d_l=\frac{\Delta \tilde{C}^{rec.}_{l}}
{\Delta \tilde{C}_{l}^{B}}\ \ ,
\ \ r_l=\langle{\frac{\tilde{C}^{rec.}_{l}-\tilde{C}^{B}_{l}}{\tilde{C}^{B}_{l}}}\rangle_{\rm{ICA}}\ \ ,
\ \ a_l=\langle{\frac{|\tilde{C}^{rec.}_{l}-\tilde{C}^{B}_{l}|}{\tilde{C}^{B}_{l}}}\rangle_{\rm{ICA}} ,
\end{equation}
meaning of which we explain now. $d_{l}$ is the ratio between the dispersion of the recovered spectra 
over 100 realizations and the quantity defined in (\ref{cosmvar}). Generally we expect this 
quantity to be larger than 1, accounting for the error introduced by the separation 
itself: a number close to 1 means that the separation procedure introduces an error 
which is negligible with respect to the input one; on the contrary, a value 
larger than 2 means that the separation error is dominating. $d_{l}$ 
is a measure of the extra uncertainty introduced by the algorithm. 

\begin{figure}
\begin{center}
\includegraphics[width=8cm]{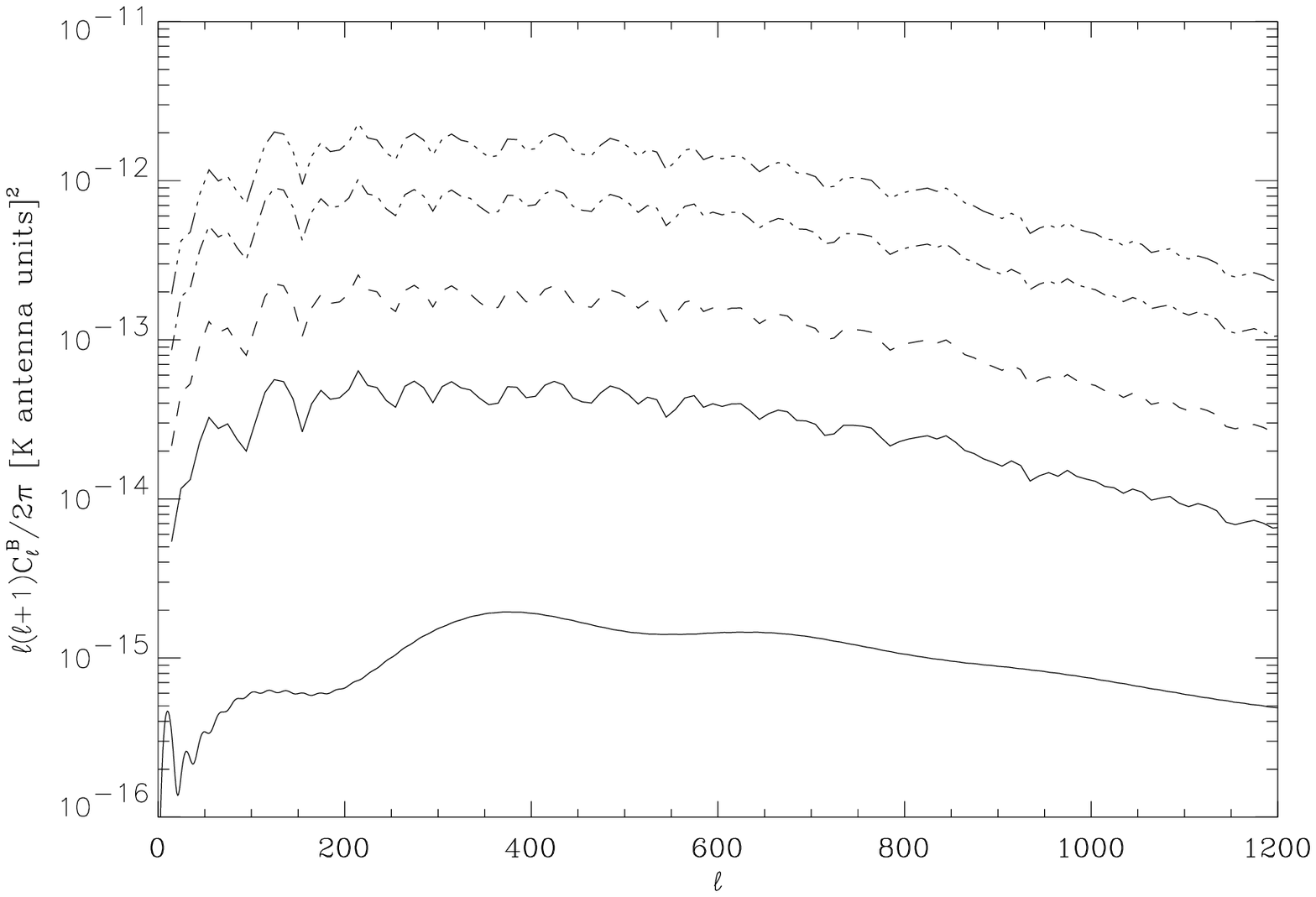}
\includegraphics[width=8cm]{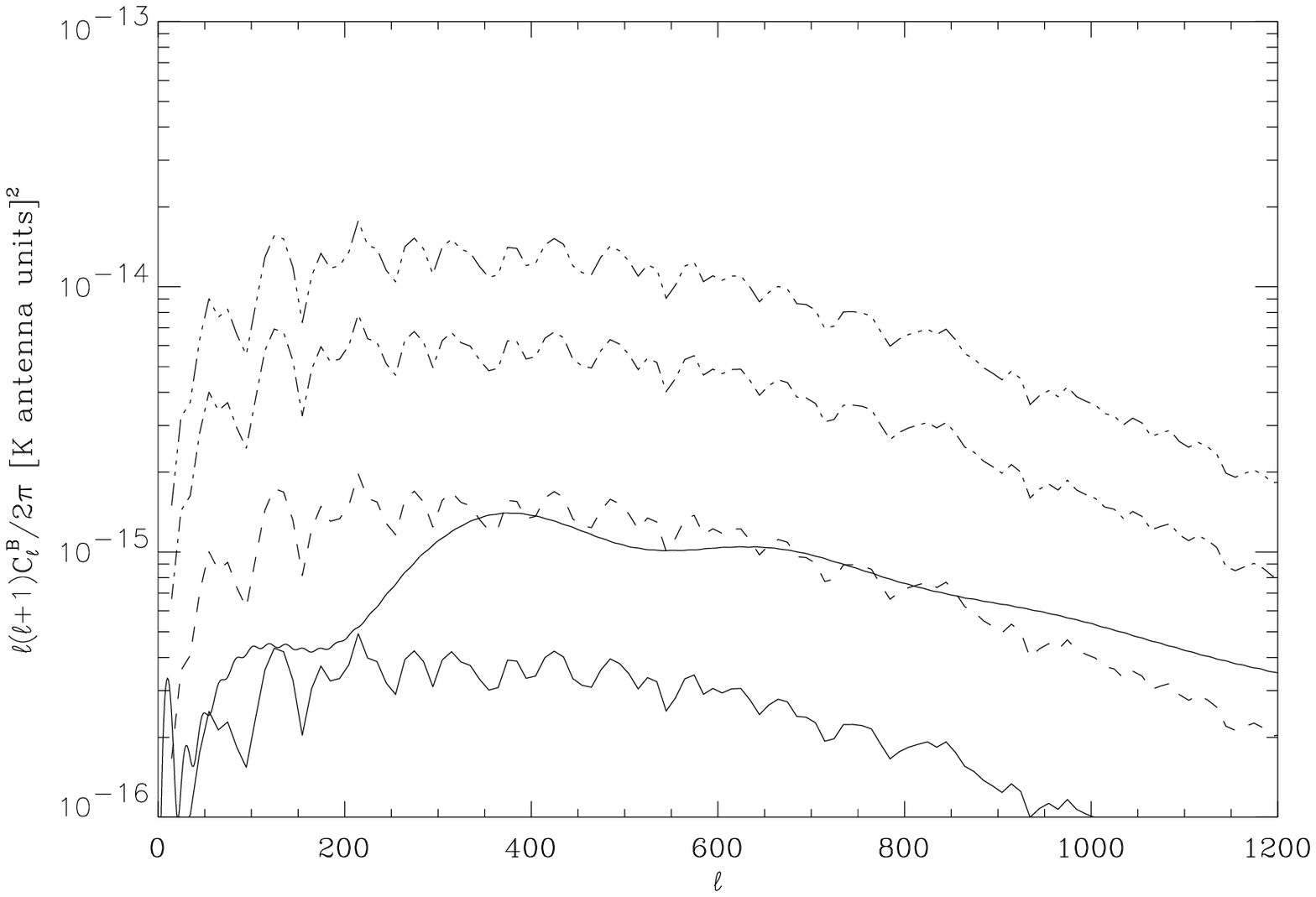}
\end{center}
\caption{Pseudo-power spectra of the synchrotron $B$ modes calculated for the 
sky region and with the amplitudes as considered in this work, at 40 (left) and 90 
GHz (right). The different curves, with raising power, correspond to the
foreground 
${\it rms}$ multiplied by 1, 2, 4, and 6, respectively. In each panel the solid 
smooth line represents the $B$ modes of the CMB.}
\label{foreground_rms_syn}
\end{figure}

\begin{figure}
\begin{center}
\includegraphics[width=8cm]{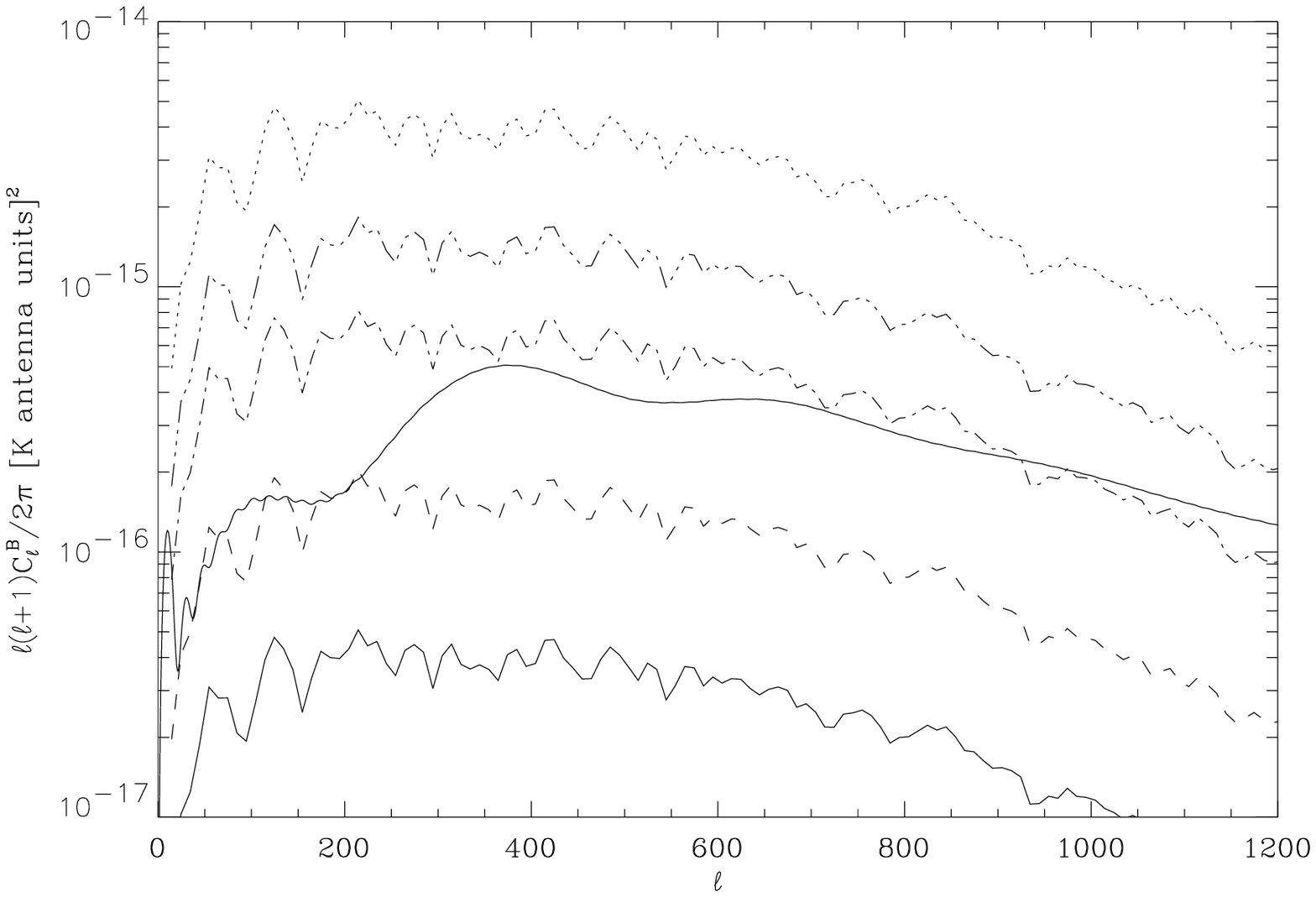}
\includegraphics[width=8cm]{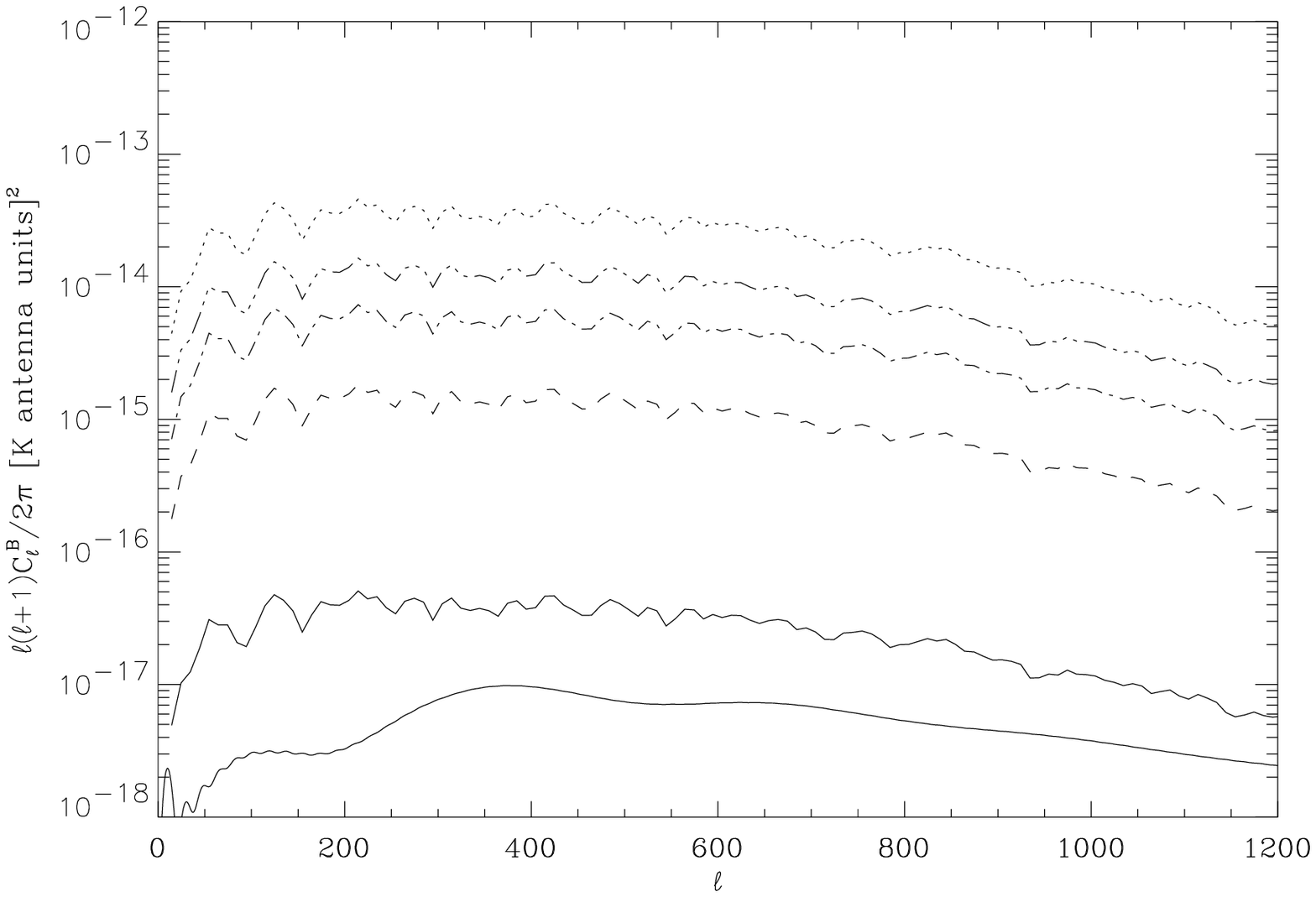}
\end{center}
\caption{Pseudo-power spectra of the dust $B$ modes calculated for the sky 
region and with the amplitudes as
considered in this work, at 150 (left) and 350 GHz (right). The different
curves, with raising power, correspond to the foreground ${\it rms}$ multiplied by 
1, 2, 4, 6 and 10, respectively. In each panel the solid smooth line 
represents the $B$ modes of the CMB.}
\label{foreground_rms_dust}
\end{figure}

On the other hand, as we see in a moment, $d_{l}$ may get closer to $1$ when 
the noise is increased, leading to the apparent paradoxical conclusion that the 
{\lightica} works better when more noise is considered. This is due to the 
fact that in some cases, the separation quality deterioration, caused by the increase of
the noise level, proceeds at the slower rate than the noise level increase itself
leading to a decrease in $d_{l}$. The $r_{l}$ and $a_{l}$ 
quantities are respectively the residual and the absolute residual 
of the recovered pseudo power spectra in the single separation, averaged over 100
realizations. These quantities give a measure of a bias of the reconstruction, 
and thus are expected to be close to zero. 
Note that part of the differences in the numerators of $r_{l}$ and $a_{l}$ 
comes from the instrumental noise; therefore, in a highly noisy configuration, 
their value can become large not because the separation fails, 
but because of the high noise itself. \\ 
Especially if checked together for each case, these quantities allow us to attempt 
to give a definition for a ``successful" separation, which is when the {\lightica} 
is able both to recover the CMB signal giving the value of $d_l$ on one side
and the values of $r_{l}$ and $a_{l}$ on the other close to, and less than 
unity, respectively. 
In tables \ref{table_r}, \ref{table_eref} and \ref{table_aeref}, we report the value 
of these quantities for some relevant multipoles as a function of the varying parameters. \\
We begin varying the noise with respect to the simulated dataset considered in the 
previous sub-section. We found out the results to be quite stable up to $S/N=1$. 
As it may be noted by looking at the first block of four rows in the tables, 
the algorithm 
performance, in terms of $d_l$, decreases mildly or remians constant, and
decreases nearly linearly with the noise amplitude in terms of $r_{l}$ and $a_{l}$. 
For of the noise larger 
than the signal, the code starts failing to reconstruct the signals, what at first shows as
a residual foreground contamination persisting in the reconstructed $B$ modes of the 
CMB, then as a failure to reach the convergence or 
to estimate a non-negative definite signal correlation matrix due to the large noise sample variance. \\
The foreground variation is realized by keeping its mean over the considered
area unchanged and increasing solely its {\it rms} by a factor 2, 4, and 6 for
synchrotron, and 2, 4, 6 and 10 for dust. In figures \ref{foreground_rms_syn} and \ref{foreground_rms_dust} 
we report the foreground $B$ modes at 40, 90 GHz and 150, 350 GHz, respectively, for 
the {\it rms} considered. For reference, we also plot the theoretical CMB pseudo-spectra. 
At 40 and 350 GHz, the contamination to the CMB is worse of course. 
Despite the high level of foreground fluctuations, the method exhibits 
again a remarkable stability or even improvement in the interval considered for 
this parameter, as it may be seen by looking at the second block of four rows 
in the tables; it starts failing only when the foreground {\it rms} is increased by a 
factor of about 6 for synchrotron, and by a factor of about 10 for dust. This
can be interpreted as due to the fact that foreground recovery is indeed easier
and more precise given a larger foreground amplitude. Indeed, for an ICA based component 
separation technique which utilizes the independence of the components to be 
recovered, the quality of the reconstruction of each of them 
depends on how well the other ones are extracted \citep{maino_etal_2002,baccigalupi_etal_2004}. \\
The last row in each table shows the effect of the variation of the sky area 
considered, while all the other parameters are kept fixed. As expected, 
things get generally better after doubling the radius of the cut, but since 
at a resolution of about 10 arcminutes a patch with $\theta_C =10^{\circ}$ has 
already a number of samples (pixels) large enough to faithfully represent the 
signal statistics, increasing $\theta_C$ doesn't improve the separation
dramatically. 
However, a wider area represents a benefit concerning the possibility of detecting the 
$B$ modes from primordial gravitational waves, as we discuss in section \ref{mtpta}. \\

As a final remark, we notice that increasing the noise amplitude causes 
the reconstruction to be less accurate, in all cases when the dust is taken into 
account. 
This observation, anticipated in Section \ref{bmraee}, is due to the fact that at 150 GHz, 
the dust emission is negligible 
with respect to CMB and noise. Indeed, as noticed in earlier works 
\citep{maino_etal_2002,baccigalupi_etal_2004}, the separation is more 
accurate when the signals are comparable in all frequency bands. This is 
supported by the fact that the performance improves or remains unaltered 
when the dust fluctuation amplitude is increased, while in the synchrotron 
case a clear degradation of the separation may be seen. \\
As a final test to evaluate the separation we study the recovered frequency scaling index
$\alpha=\log{[s(\nu_{2})/s(\nu_{1}))]}/\log{(\nu_{2}/\nu_{1})}$ of the different 
output components $s$, computed through the ratios between column elements in the inverse 
of the separation matrix \citep{maino_etal_2002}. In all the cases we studied, 
this quantity resulted to be close to the theoretical one, with dispersions 
$\Delta\alpha$ increasing roughly linearly with foreground amplitude and noise; 
an exception is still represented by the dust case, when the dust {\it rms} is 
increased: as explained above, increasing the dust {\it rms} induces an improvement in the 
reconstruction, which appears also in the recovery of $\alpha$. The relative
dispersion $\Delta\alpha/\alpha$, evaluated over the 100 Monte Carlo realizations
both for CMB and foregrounds, is shown in table \ref{table_freq_sca}.
\begin{table}
\begin{center}
\begin{tabular}{|c c c c c c c c c|}
\hline
 $S/N$ & Fore. & Cut  & $d_{l=100}$ &  $d_{l=400}$ & $d_{l=950}$ & $d_{l=100}$ &  $d_{l=400}$ & $d_{l=950}$   \\
 & Ampl. & Radius&  Sync. & Sync. &  Sync. &  Dust & Dust&  Dust\\
\hline
\hline
 $\infty$ & 1.00 & 10 & 1.16 &  1.69 &  1.31 &  2.08& 1.99& 1.74  \\
\hline
 $2.00$ & 1.00 & 10 & 1.44  & 1.55 &1.18 & 1.81& 1.56& 1.31    \\
\hline
 $1.50$ & 1.00 & 10 & 1.38 & 1.42&1.07&  2.24  &1.56&1.31 \\
\hline
 $1.00$ & 1.00 & 10 & 1.88 & 1.72& 1.38 &  3.39&1.56& 1.21\\
\hline
\hline
$2.00$ & 2.00 & 10 & 1.60 &  1.70& 1.33 & 1.66  & 1.64&  1.34\\
\hline
$2.00$ & 4.00 & 10 & 1.62  & 2.03& 1.58 &1.63& 1.69& 1.33 \\
\hline
$2.00$ & 6.00 & 10 & 1.78  & 2.39& 1.81 &1.51& 1.69& 1.32\\
\hline
$2.00$ & 10.00 & 10 & -  & -& -& 2.00& 2.04&  1.50 \\
\hline
\hline
$2.00$ & 1.00 & 20 & 1.28 &  1.53&  1.03 & 1.75 & 1.41& 1.24 \\
\hline
\end{tabular} 
\end{center}
\caption{Relative extra uncertainty, $d_l,$ evaluated for the reconstructed 
$B$ mode power spectrum of the CMB. The results are given for three different 
values of the multipole $l$ and for multiple choices of the sky and noise 
parameters as listed in the table.}
\label{table_r}
\end{table}

\begin{table}
\begin{center}
\begin{tabular}{|c c c c c c c c c|}
\hline
 $S/N$ & Fore. & Cut  & $r_{l=100}$  & $r_{l=400}$  & $r_{l=950}$ & $r_{l=100}$  & $r_{l=400}$  & $r_{l=950}$   \\
 & Ampl. & Radius&  Sync. &  Sync.&   Sync. &  Dust&  Dust &  Dust \\
\hline
\hline
 $\infty$ & 1.00 & 10 & 0.02 & 0.05&  0.05 & 0.10&  0.05& 0.05\\
\hline
 $2.00$ & 1.00 & 10 &  0.01 &  0.06&  0.08 & 0.11&  0.06&  0.07\\
\hline
 $1.50$ & 1.00 & 10 &  0.02& 0.06&0.10& 0.13& 0.06& 0.17 \\
\hline
 $1.00$ & 1.00 & 10 & 0.02 & 0.08&0.22& 0.22& 0.08& 0.20\\
\hline
\hline
$2.00$ & 2.00 & 10 &  0.01&  0.06& 0.08& 0.08  & 0.05& 0.05\\
\hline
$2.00$ & 4.00 & 10 & 0.02  & 0.08&0.14&0.08 & 0.06& 0.06\\
\hline
$2.00$ & 6.00 & 10 &  0.02&  0.08&0.14& 0.08& 0.06& 0.03 \\
\hline
$2.00$ & 10.00 & 10 & -  & -& -& 0.08& 0.08&  0.05 \\
\hline
\hline
$2.00$ & 1.00 & 20 & 0.02 & 0.04& 0.03&0.09  &0.05  &0.03  \\
\hline
\end{tabular} 
\end{center}
\caption{Residuals of the CMB pseudo $B$ modes recovered against synchrotron 
and dust, averaged over 100 realizations of noise and CMB.}
\label{table_eref}
\end{table}
\begin{table}
\begin{center}
\begin{tabular}{|c c c c c c c c c|}
\hline
 $S/N$ & Fore. & Cut  & $a_{l=100}$ & $a_{l=400}$ & $a_{l=950}$ & $a_{l=100}$ & $a_{l=400}$ & $a_{l=950}$  \\
 & Ampl. & Radius&  Sync.&  Sync. &  Sync. &  Dust &  Dust&   Dust  \\
\hline
\hline
 $\infty$ & 1.00 & 10 & 0.12& 0.06& 0.06&  0.14 & 0.06&  0.05 \\
\hline
 $2.00$ & 1.00 & 10 & 0.12& 0.07& 0.11& 0.15 & 0.07 &  0.14 \\
\hline
 $1.50$ & 1.00 & 10 & 0.12& 0.08& 0.17&0.17 & 0.07&  0.17\\
\hline
 $1.00$ & 1.00 & 10 & 0.12& 0.11& 0.46&  0.22 & 0.10&  0.29\\
\hline
\hline
$2.00$ & 2.00 & 10 & 0.09& 0.07& 0.14&   0.13  &0.06  &  0.13\\
\hline
$2.00$ & 4.00 & 10 & 0.10& 0.11& 0.18&  0.13&  0.07 & 0.14\\
\hline
$2.00$ & 6.00 & 10 & 0.10& 0.11& 0.19&   0.13& 0.08& 0.14 \\
\hline
$2.00$ & 10.00 & 10 & -  & -& -& 0.13& 0.09& 0.13  \\
\hline
\hline
$2.00$ & 1.00 & 20 & 0.09 & 0.07& 0.09& 0.09 & 0.05 & 0.08 \\
\hline
\end{tabular} 
\end{center}
\caption{Absolute value of residuals of the CMB pseudo $B$ modes recovered 
against synchrotron  and dust, averaged over 100 realizations of noise and CMB.}
\label{table_aeref}
\end{table}
\begin{table}
\begin{center}
\begin{tabular}{|c c c c c c c|}
\hline
 $S/N$ & Fore. & Cut  & CMB vs. Sync.& Synchrotron & CMB vs. Dust & Dust  \\
 & Ampl. & Radius \\
\hline
\hline
 $\infty$ & 1.00 & 10 & 0.01& 0.03& 0.72& 0.81\\ 
\hline
 $2.00$ & 1.00 & 10 & 0.05& 0.07& 0.60& 0.89\\ 
\hline
 $1.50$ & 1.00 & 10 & 0.14& 0.15& 0.75& 1.09\\ 
\hline
 $1.00$ & 1.00 & 10 & 0.28& 0.35& 1.17& 1.97\\ 
\hline
\hline
$2.00$ & 2.00 & 10 & 0.14& 0.14& 0.63& 0.69\\ 
\hline
$2.00$ & 4.00 & 10 & 0.22& 0.23& 0.85& 0.56\\
\hline
$2.00$ & 6.00 & 10 &  0.35& 0.36& 1.49& 0.53\\
\hline
$2.00$ & 10.00 & 10 & -  & -& 3.04& 0.46 \\
\hline
\hline
$2.00$ & 1.00 & 20 & $<0.01$ & 0.02& 0.45& 0.73 \\
\hline
$1.00$ & 1.00 & 20 & 0.18 & 0.20& - & - \\
\hline
\end{tabular} 
\end{center}
\caption{Relative dispersions $\Delta\alpha/\alpha$ around the expected 
values of the  frequency spectral indices, for both CMB and foregrounds.}
\label{table_freq_sca}
\end{table}

\section{Measuring the primordial tensor amplitude}
\label{mtpta}

As we stressed already, one of the most important goals of the forthcoming CMB 
polarization experiments is the measure of the ratio $r$ between the primordial 
amplitudes of tensor and scalar cosmological perturbations, i.e. gravity 
waves and density fluctuations. The most relevant question in this context 
is how small that ratio can be in order to be detected when foregrounds are taken 
into account, and in particular what this minimal detectable value is when 
the CMB background is separated 
from the foregrounds with the technique considered here. As we stressed in section 
\ref{sm}, the foreground simulations are still too uncertain to push the analysis 
toward a complete cosmological parameter estimation pipeline and address this
question comprehensively.
Nevertheless, given the importance of this topic, we present in this section
some general though preliminary remarks  and we illustrate them with some
examples.

For our purpose here the most relevant result of the previous sections
is the observation that the {\fastica}-based separation yields errors which 
are comparable to those from cosmic variance and noise, for the model 
with $r=0.1$. Therefore, in such a case, as far as these simulations 
are concerned, we should be able to detect the tensor contribution in the 
presence of foregrounds when the latter are treated with ICA, with a 
confidence close to the one achievable without foregrounds.

\begin{figure}
\begin{center}
\includegraphics[width=8cm]{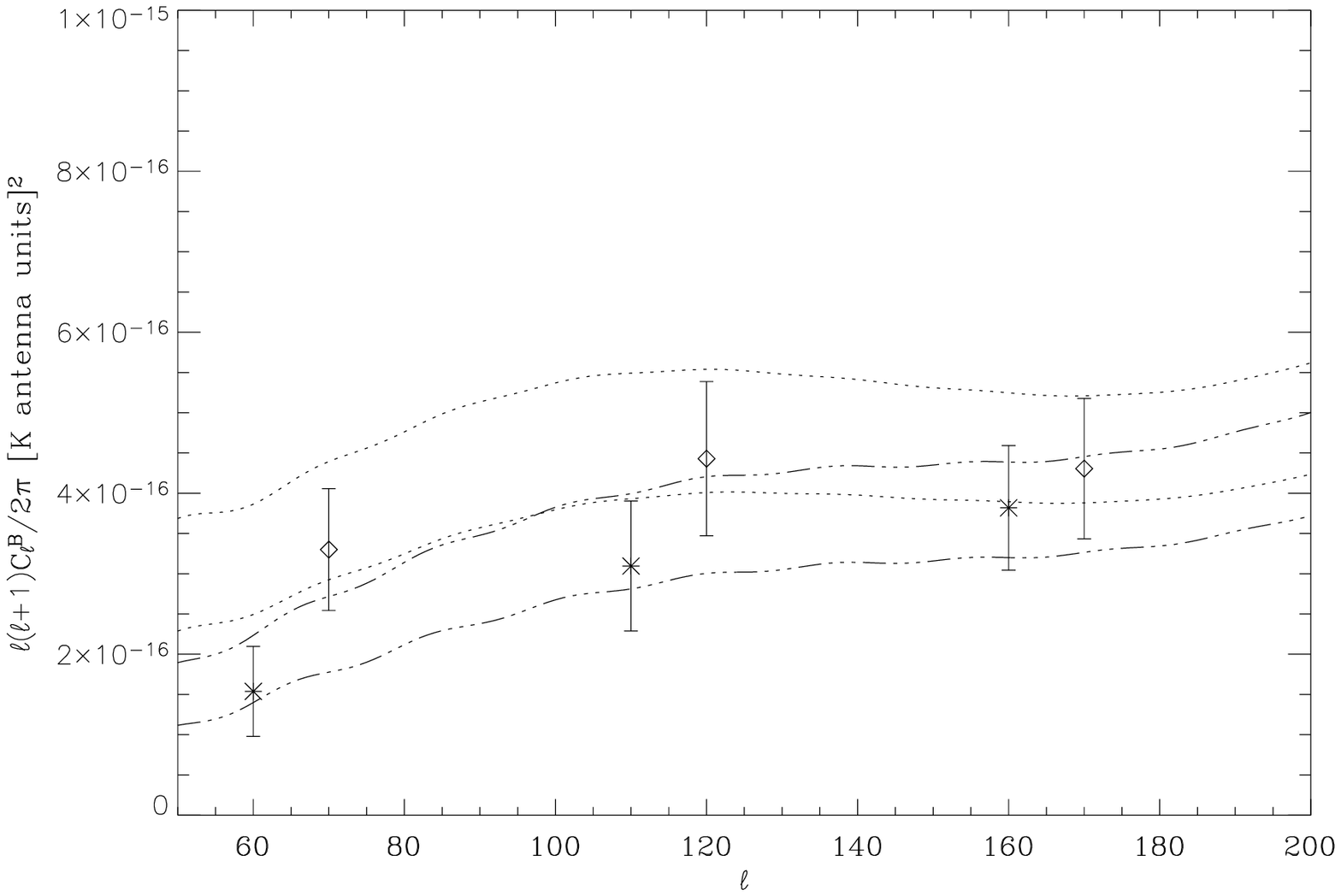}
\includegraphics[width=8cm]{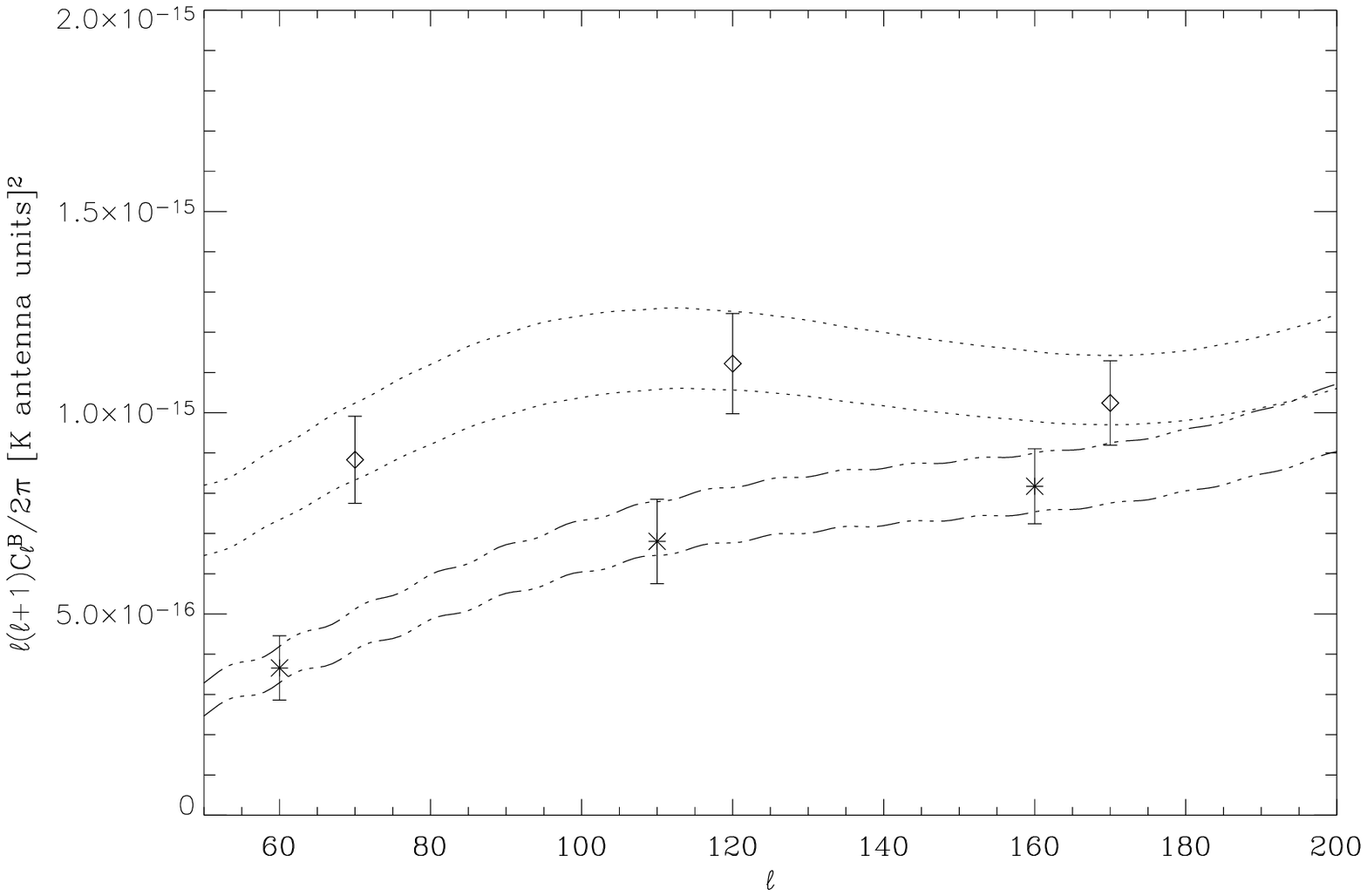}
\end{center}
\caption{Comparison of the recovered pseudo-$B$ modes in two different 
cosmological models with $r=0$ (dashed-dotted lines and asterisks) and 
$r=0.1$ (dotted lines and diamonds shifted by $\Delta l=10$ for clarity), in
the low frequency combination 
with $S/N=2$. Left panel refers to $\theta_C=10^{\circ}$ , while right
panel to $\theta_C=20^{\circ}$. Regions between the lines are the dispersion
coming from cosmic variance and noise.}

\label{model_comp}
\end{figure}

To illustrate this issue, we compare the pseudo $B$ mode recovery in our fiducial 
model with $r=0.1$ with one in which the tensors are absent, $r=0$. We focus 
on the spectral region where primordial tensors are most relevant, e.g. $l\simeq 100$. 
Of course, as a result of the leakage of $E$ modes into $B$ due to the limited sky 
coverage, also in the latter case the amplitude of $E$ 
modes on these scales also matters. We address this issue to some extent
by considering the results for each value of $r$ obtained for sky areas with different aperture. 
In figure \ref{model_comp} we plot the recovered pseudo $B$ modes in these two cases, 
zooming on the relevant range of multipoles. Two different sizes for the sky 
cut are considered 
here; the left and right panel refer to $\theta_C=10^{\circ}$ and $\theta_C=20^{\circ}$, 
respectively. In both panels the higher amplitude spectrum represents the model with 
$r=0.1$. As we quoted above, we see that in the entire
interval, the separation error is comparable to the cosmic variance, 
in particular in the $l$-range, where the $B$ modes from tensors have their main impact, 
i.e. $l\simeq 100$. 
The central role of the leakage from E modes is evident. When the area
gets smaller and the pollution consequently larger, the detection of $B$ modes becomes
harder. 
Points on the plot are the pseudo $B$ modes recovered by the
code against the synchrotron template, in the case with $S/N=2$, for both the
models (asterisks for $r=0$ and diamonds $r=0.1$). 

Looking at the left panel of the figure, it is clear that, even if the
algorithm did not introduce any extra uncertainty, we could not make 
any claim on primordial $B$ modes detection with sufficient confidence,
mostly beacuse the leakage is already too high for the 10 deg cut. 
The situation gets better for the $\theta_C=20^{\circ}$ case (the right
panel), where the two models seem to be distinguishable statistically by our method.
For the latter case, more quantitative results are showed in table \ref{stat_model_comp}. 
To define whether or not the algorithm is able to distinguish between the two
models, we compare the statistics of the recovered power spectra. In the
second column, we report the percentage of recovered power spectra (for
$r=0$) that fall inside the $95\%$ confidence region for $r=0.1$ as calculated
directly from the statistics of the recovered power spectra themself. In this way, 
we obtain the probability for ICA to give a false detection of tensor 
contribution. Viceversa, the
fourth column shows the probability to miss a true presence of primordial B
modes, since it reports the percentage of recovered power spectra for
$r=0.1$ that fall inside the $95\%$ confidence region for $r=0$. Third and
fifth column give the same probabilities but computed for simulated, CMB-only
spectra and thus not requiring any further processing. These columns provide an 
idea of the best achievable levels.

We thus conclude that, given the available foreground 
simulations, {\fastica} eliminates the 
foregrounds in the two cases of r we consider, with
a precision sufficient to make them distinguishable even
with the suboptimal pseudo-BB estimator.
As already stressed above, this is 
due to the fact that the separation process induces an error comparable 
to those coming from cosmic variance and noise. 

Further analysis on the recovering of the
true $B$ modes and, mostly important, on the minimum value of $r$ that 
can be detected with this technique, could be performed (for problems 
relevant to this issue and what can be expected for the considered cases, 
see e.g. \cite{lewis_2002, lewis_2003} and reference therein).
However they would not only go beyond the intended aim of this work, 
which is to show the capability of ICA in a situation close to what 
we expect for the forthcoming CMB polarization experiments, but could also 
prove to be misleading given the substantial uncertainties that 
still affect the foreground simulations and because of the absence of 
systematics in this simulation.  We actually plan to address these issues 
in a future work, probably exploiting the new {\it pure} pseudo power 
spectrum estimator from \cite{smith_2005}, in
a new, more realistic simulated environment.

  \begin{table}
  \begin{center}
  \begin{tabular}{|c c c c c |}
  \hline
     & $\tilde{C}^{out}_{r=0}$  vs $\tilde{C}^{out}_{r=0.1}$
     &$\tilde{C}^{in}_{r=0}$  vs $\tilde{C}^{in}_{r=0.1}$  &
     $\tilde{C}^{out}_{r=0.1}$ vs $\tilde{C}^{out}_{r=0}$ &
     $\tilde{C}^{in}_{r=0.1}$  vs $\tilde{C}^{in}_{r=0}$ \\
  \hline
  \hline
   $\ell=70$ & $0\%$ & $0\%$ & $0\%$ & $0\%$ \\
  \hline
   $\ell=120$ &  $0\%$ & $0\%$& $5\%$ & $2\%$ \\
  \hline
   $\ell=170$ &  $50\%$ &  $43\%$& $53\%$ & $46\%$ \\
  \hline
  \hline

  \end{tabular} 
  \end{center}
  \caption{This table reports the capability of the algorithm to distinguish
     between the two models with and without the gravity wave content
     for the $\theta_C=20^{\circ}$ case. The second column shows the 
     probability (see text for more details)
     a spurious detection of the tensor contribution, while the fourth column 
     shows the probability of a false non-detection.
     The other two columns report the same probabilities but derived
     in the ideal, CMB-only, foreground-free case.}
  \label{stat_model_comp}
  \end{table}

\section{Discussion}
\label{d}

In this work we discuss the performance of the component separation technique
based on the Independent Component Analysis (ICA) as applied to the small patches 
of the polarized sky observed in the microwave band. We focus on the recovery 
of the Cosmic Microwave Background (CMB) signal out of the contamination due to 
the diffuse foreground emission from the Galaxy and perform a Monte Carlo analysis 
to estimate the resulting errors. These computations employ {\lightica} -- a newly 
developed parallel implementation of the {\fastica} code 
\citep{hyvarinen_1999,maino_etal_2002}. The sky region considered here is and has been,
the target of several CMB experiments due to its known low 
foreground emission in total intensity. Nevertheless, the current 
knowledge of the foreground emissions indicates that their contaminations to 
the CMB emission in polarization may be non-negligible in any region of the 
sky, and at any frequency considered, including sky areas which appear nearly free of the
foregrounds in total intensity.
This observation is particularly relevant for the component of 
the CMB signal due to the primordial 
gravitational waves and lensing mechanisms ($B$ modes); the latter 
signal is extremely relevant in modern cosmology as it may demonstrate 
the existence of gravitational waves of cosmological origin, as well 
as reveal crucial clues about the structure formation in the universe. \\
The idea behind the ICA is based on the 
assumption that the background and foreground statistics are independent, 
requiring no other prior on the signals to recover, on their 
pattern or frequency scaling. All the available data and simulations 
on the Galactic foreground emission indicate that the CMB and Galactic 
emission are highly statistical independent; indeed, the CMB is known to 
have a statistics which is close to the Gaussianity, while the Galaxy is 
known to be highly non-Gaussian. This occurrence, together with the high 
level of detail in the present CMB data, reaching the arcminute scale, 
allow the ICA technique to recover the CMB pattern extremely close to the 
actual one. \\
We consider two sets of two frequency bands 
where the foreground contamination is given by the synchrotron and 
the thermal dust, respectively, and we assume a Gaussian and uniform
noise distribution with amplitude comparable to the total CMB polarized 
emission. We quantify the quality of the reconstruction by comparing 
the input and output angular power spectra on the sky fraction which we consider 
(pseudo power spectra). Due to its parallelized structure, the {\lightica} 
allows to evaluate the error in the separation process on each relevant 
angular scale, via Monte Carlo simulations sampling some of the 
most relevant degrees of freedom entering in the dataset simulation. 
We identify the error induced by the separation process on the CMB 
reconstruction and show that is comparable or lower than the uncertainty given 
by the instrumental noise and cosmic variance. We remark that in terms of 
pseudo-$B$ modes, this is 
achieved in presence of a foreground contamination which may be several times 
stronger than the cosmological signal, which has been simulated accordingly 
to the current polarized foreground models. Then, we evaluate the stability 
of these results against variation of the noise and foreground fluctuation 
amplitude, as well as the sky area covered and the abundance of primordial 
gravitational waves. 
We find considerable intervals of these parameters where the results
do not change substantially. In particular, the claim that 
the error induced by the reconstruction is of the order of the cosmic 
plus noise sample variance, remains true in all these cases. Moreover, 
the outputs exhibit a remarkable stability for a large foreground fluctuation 
amplitude, several times larger than predicted by the current foreground models; 
this may be due to a compensation between the degradation which would be 
induced by the large foreground signal, and the fact that the latter 
is better extracted as it gets larger over the noise. 
Then, we found that increasing the area covered yields an improvement 
of the separation performance, not too large since the smaller cut already
contains enough statistical information to provide good convergence. By 
comparing the results with smaller and higher areas when the primordial 
tensor to scalar ratio is $r=0.1$ and $r=0$, we were able to check the 
benefit of having a larger 
area for measuring $r$, due to the weaker leakage from $E$ modes. 
We were able to verify that for an area large as $20\times 20$
squared degrees, the cases with r=0.1 and 0 are distinguishable
even at the level of the suboptimal pseudo B mode power spectra at $l=100$, 
after ICA has removed a foreground contamination which may be several times
larger than the cosmological signal.
We stress that this claim needs more investigation in the future 
due to the foreground uncertainties and 
absence of systematics, as we explain below. 

We did not specialize our treatment for any specific 
experimental setup, but rather focused on a handful of important 
but quite general and basic parameters. 
This is because our aim is to show that if the foreground 
contamination to the $B$ modes of the CMB turns out to be 
consistent as the current foreground models predict, then an 
ICA based component separation 
may be required and adequate to achieve the detection of the cosmological 
signal. It is still premature to attempt a quantitative estimation of 
the ICA nominal performance taking into account the uncertainty in the 
foreground signal, in general or for a given experiment. Indeed, as we 
explain below the available information on the Galactic polarized signal is still 
too modest to assess properly its statistical distribution.
Nevertheless, it is worthwhile to compare frequencies, angular resolution and 
noise amplitude 
of the planned and ongoing CMB polarization observations, with our assumptions. 
We thus note that the configuration 
at high frequencies, for angular resolution and noise amplitude, is close 
to the BOOMERanG case \citep{montroy_etal_2005}. Our high frequency 
combinations are also close to those adopted by EBEx \citep{oxley_etal_2004} 
and QUAD \citep{bowden_etal_2004}, while 40 and 90 GHz setup is adopted by 
QUIET. These two experiments have also an angular resolution close 
to what  we assumed herein, however, given that
they explicitely aim at the $B$ mode detection, their 
nominal noise amplitude is markedly lower with respect to what we 
considered here and what was set to be comparable to the CMB 
${\it rms}$ in $Q$ or $U$ maps and hence to that of the $E$ mode.\\ 
It is useful to recall here the limitations of the present analysis. First of 
all, a definitive assessment of the performance of ICA or any other component 
separation technique should come when the foregrounds simulations are put on
a firmer ground. At the present the models are built upon observations in the 
radio or far infrared bands 
as well as the WMAP three year data \citep{page_etal_2006}, used to
constraint the polarization angle pattern, while the polarization fraction with respect 
to total intensity is often inferred from large scale data only. Moreover, 
the effects of extra-Galactic point sources were not considered here.
The basic treatment of them in an 
experiment aiming at the detection of the diffuse signal is the identification and 
removal of the brightest ones, usually at 5$\sigma$ from the {\it rms} of the diffuse 
signals plus noise; the remaining unresolved ones yield an angular 
power spectrum also rising approximately as $l^{2}$. More sophisticated techniques involve 
the use of wavelets to identify and remove also fainter sources \citep{vielva_etal_2003}, 
which have been proved so far promising for the identification and removal of all 
sources down to a flux comparable with the one from the diffuse signals plus 
noise. Actually, according to the latest models \citep{tucci_etal_2004}, 
the angular power spectrum in $B$ from extra-Galactic radio sources, 
where only the brightest have been removed, is actually comparable with the 
noise we consider in this work at 40 GHz, being subdominant at higher frequencies. 
An analogous situation might occur at 350 GHz, from the contribution to infrared 
sources if they possess an high polarization ratio, say one percent or more, 
although those are very poorly known. Our separation results were 
quite stable also for a lower signal to noise ratio, which seems to 
indicate that if the {\it rms} of the unresolved sources in the maps is 
estimated correctly, then the ICA should be relatively insensitive to 
them. This claim is certainly premature due to the foreground uncertainties, 
affecting extra-Galactic sources together with the diffuse Galactic signal, making 
difficult an accurate estimation of the contribution from unresolved sources to 
the noise, and possibly requiring a deeper source removal with appropriate
techniques. Another important aspect concerns the simulation of the CMB
itself. The latter is evaluated assuming a Gaussian statistics, while the
lensing distortion, which is responsible for a large portion of the signal 
in the $B$ polarization modes, causes a non-Gaussian distortion. The latter 
is caused by the correlation of different cosmological scales 
induced by the lensing mechanism. Numerical machineries to lens a Gaussian CMB 
realization are becoming available, see \cite{lewis_2005}. Although this issue 
is still under investigation, in particular for the effect on pseudo-$C_{l}$s 
on a limited patch of the sky, it has to be considered in future works on the 
present subject; in a MonteCarlo pipeline for evaluating the errors of the CMB 
reconstruction in the present and other cases, one should use the proper 
CMB templates, i.e. obtained by varying the primordial Gaussian realization 
as well as its non-Gaussian lensing distortion.\\
It is also important to stress the ICA fundamental hypotheses and limitations 
at the present level of the code architecture. 
Together with the statistical independence, which is likely to be very well 
verified for the CMB and the Galaxy, a fundamental limitation is represented 
by the assumption that the signals scale rigidly in frequency, which means 
that the spectral index is spatially independent. 
This is not verified on the whole sky, as both the energy distribution 
of free electrons, as well as the thermal dust temperature, 
exhibit fluctuations. On the other hand, so far there is no evidence 
that the synchrotron spectral index actually vary substantially on 
the angular scales considered in experiments observing a limited 
fraction of the sky, which is usually comparable to a percent of 
the sky. In addition, in the real space the ICA requires to work 
at the same angular resolution at all frequencies, which is not the case 
for most of the operating of planned observations. This makes necessary to 
work at the lowest resolution, or an approach in the harmonic space should 
be implemented. 
The ICA technique has also degrees of freedom which were not fully 
exploited yet. First, the neg-entropy approximation which is being used 
is still the original one, which was introduced in other context, with very 
different purposes from the present, cosmological applications. The question 
whether or not a more appropriate neg-entropy approximation exists for the problem 
at hand is still open. Moreover, the possibility to use priors on the mixing matrix 
and/or foregrounds should be taken into account, as it was done by \citet{maino_etal_2003} 
for the first time. In polarization, where foregrounds are most uncertain, 
a possible constraint could come from the black-body frequency scaling
for the CMB,  which was not exploited so far. \\
Ultimately, all our results have been obtained in absence of instrumental 
systematics. The stability of the performance in these nominal conditions has 
to be quantified and verified in presence of the most important systematics 
effects, such as non-uniform and non-gaussian noise, beam asymmetry, etc.

However, we believe that the excellent performance shown here justifies 
the interest and effort toward the implementation of techniques based on the 
Independent Component Analysis in real experimental conditions. Even if only 
a fraction of their capability remains in a real experiment application, that 
might be crucial in order to measure the actual pattern of the B-mode component 
of the CMB polarized emission. 

\section*{Acknowledgments}

Carlo Baccigalupi is grateful to George F. Smoot for several 
useful discussions. Some of the results in this paper have been derived using the 
Hierarchical Equal Area Latitude Pixelization of the sphere (HEALPix, \cite{gorski_etal_2005}). 
The theoretical power spectra were calculated using
the {\sc cmbfast} software by \citet{seljak_zaldarriaga_1996}.
We acknowledge the use of National Energy Research Scientific Computing
Center computing resources. This research 
was supported in part by the NASA LTSA grant NNG04GC90G.

\end{document}